\documentclass[aps,twocolumn,preprintnumbers]{revtex4-1}
\pdfoutput=1
\usepackage{bm,amsmath,amssymb,amsfonts,slashed,array,graphicx,soul}
\usepackage{mathrsfs}
\usepackage{xcolor}
\usepackage{hyperref}
\usepackage{cancel}
\usepackage[normalem]{ulem}
\usepackage{slashed}            
\usepackage{tikz}
\usetikzlibrary{arrows,shapes}
\usetikzlibrary{trees}
\usetikzlibrary{matrix,arrows} 				
\usetikzlibrary{positioning}				
\usetikzlibrary{calc,through}				
\usetikzlibrary{decorations.pathreplacing}  
\usetikzlibrary{decorations.pathmorphing}	
\usetikzlibrary{decorations.markings}
\usetikzlibrary{snakes}
\usepackage[normalem]{ulem}

\def\mht{m_{\tilde{h}}}

\def\alam{\alpha_\lambda}
\def\lsim{\mathrel{\rlap{\lower4pt\hbox{\hskip1pt$\sim$}}
    \raise1pt\hbox{$<$}}}
\def\gsim{\mathrel{\rlap{\lower4pt\hbox{\hskip1pt$\sim$}}
    \raise1pt\hbox{$>$}}}

\begin{document}
\title{Electroweak-Charged Bound States as LHC Probes of Hidden Forces}
\author{Lingfeng Li,$^{1}$ Ennio Salvioni,$^{2}$ Yuhsin Tsai,$^{3}$ and Rui Zheng$^{1\;}$}
\email[Email: ]{llfli@ucdavis.edu}\email{ennio.salvioni@tum.de}\email{yhtsai@umd.edu}\email{ruizh@ucdavis.edu}
\affiliation{$^{1}$Department of Physics, University of California, Davis, Davis, California 95616, USA\\
$^{2}$Physics Department, Technical University of Munich, 85748 Garching, Germany\\
$^{3}$Maryland Center for Fundamental Physics, Department of Physics, University of Maryland, College Park, Maryland 20742, USA}
\preprint{TUM-HEP-1103-17, UMD-PP-017-032}
\begin{abstract}
We explore the LHC reach on beyond-the-Standard Model (BSM) particles $X$ associated with a new strong force in a hidden sector. We focus on the motivated scenario where the SM and hidden sectors are connected by fermionic mediators $\psi^{+, 0}$ that carry SM electroweak charges. The most promising signal is the Drell-Yan production of a $\psi^\pm \bar{\psi}^0$ pair, which forms an electrically charged vector bound state $\Upsilon^\pm$ due to the hidden force and later undergoes resonant annihilation into $W^\pm X$. We analyze this final state in detail in the cases where $X$ is a real scalar $\phi$ that decays to $b\bar{b}$, or a dark photon $\gamma_d$ that decays to dileptons. For prompt $X$ decays, we show that the corresponding signatures can be efficiently probed by extending the existing ATLAS and CMS diboson searches to include heavy resonance decays into BSM particles. For long-lived $X$, we propose new searches where the requirement of a prompt hard lepton originating from the $W$ boson ensures triggering and essentially removes any SM backgrounds. To illustrate the potential of our results, we interpret them within two explicit models that contain strong hidden forces and electroweak-charged mediators, namely $\lambda$-supersymmetry (SUSY) and non-SUSY ultraviolet extensions of the Twin Higgs model. The resonant nature of the signals allows for the reconstruction of the mass of both $\Upsilon^\pm$ and $X$, thus providing a wealth of information about the hidden sector. 
\end{abstract}
\maketitle

\section{Introduction}\label{sec:intro}
New hidden particles that couple weakly to the Standard Model (SM), but interact strongly with other beyond-the-SM (BSM) states, play important roles in theories addressing the electroweak hierarchy problem, such as neutral naturalness \cite{Chacko:2005pe,Burdman:2006tz} and natural supersymmetry (SUSY) \cite{Batra:2003nj,Maloney:2004rc,Barbieri:2006bg}, as well as in models that explain cosmological anomalies \cite{Tulin:2013teo,Hochberg:2014dra,Kaplinghat:2015aga}. Examples of such particles, which in this paper are called {\it hidden force carriers}, include hadrons bound by a new confining interaction, or the physical excitations associated with a new scalar or vector force. 

Testing the existence of hidden force carriers is an important task of the Large Hadron Collider (LHC). Since these typically have small couplings to the SM sector, however, their direct production is very suppressed. Nevertheless, in many motivated BSM scenarios other new particles exist, charged under at least some of the SM symmetries, that can serve as {\it mediators} to access the hidden force carrier at the LHC. In this paper we focus on the challenging, but motivated, case where the mediators, labeled $\psi$, have SM electroweak (and not color) charges. Once a $\psi \bar{\psi}$ pair is produced via the electroweak interactions, it can form a bound state held together by the hidden force. Since the hidden force carrier $X$ has a large coupling to the mediators, it is produced with sizable probability in the ensuing bound state annihilation, possibly in association with other SM object(s) to ensure electroweak charge conservation. $X$ can then decay through its small coupling to SM particles, yielding either prompt or displaced signatures in the LHC detectors.
\begin{figure}
\begin{center}
\includegraphics[width=6cm]{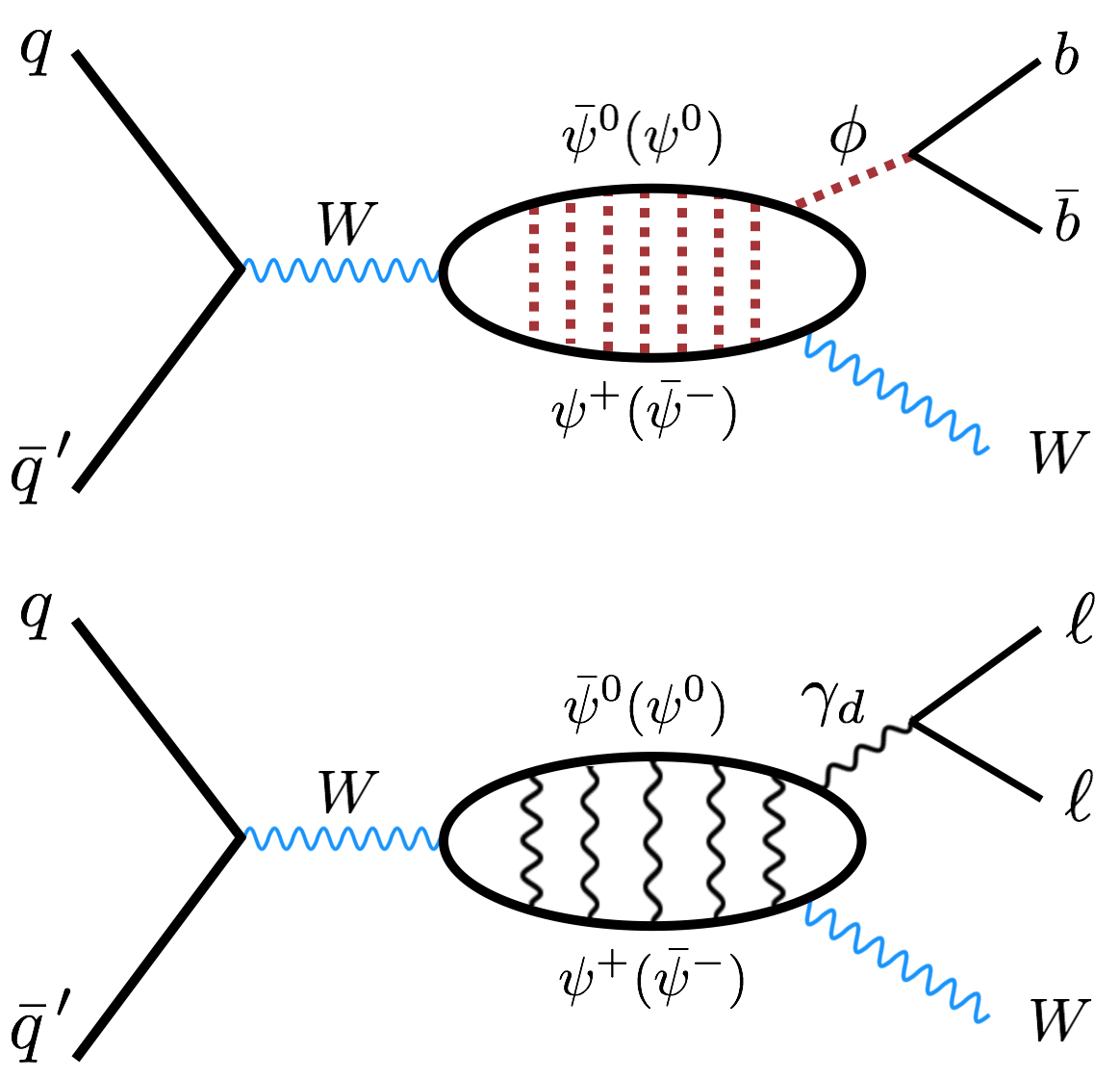}
\end{center}
\caption{The collider processes studied in this paper. Here $\psi^{+,0}$ are the {\it mediators}, new particles that carry SM electroweak (but not color) charge, which we take to be vector-like fermions. Their Drell-Yan pair production leads to the formation of electroweak-charged bound states due to a hidden force. The annihilation of the electrically charged bound state $\Upsilon^\pm$ produces a $W^\pm X$ pair, where $X$ is the {\it hidden force carrier}. We take $X$ to be either a real scalar $\phi$, which decays back to the SM via mass mixing with the SM-like Higgs, or a dark photon $\gamma_d$ that decays via kinetic mixing with the SM photon. The $\phi \to b\bar{b}$ and $\gamma_d \to \ell \ell$ decays are selected, which can be either prompt or displaced on collider timescales.}
\label{fig:feynman}
\end{figure} 

For concreteness, in this paper we consider the cases where the hidden force carrier is either a real scalar or a dark photon, \mbox{$X = \phi, \gamma_d$}, while the mediators are a pair of vector-like fermions $\psi^{+, 0}$, with the superscript indicating the SM electric charge. The relevant LHC processes are shown in Fig.~\ref{fig:feynman}: A $\psi^+ \bar{\psi}^0$ (or $\psi^0 \bar{\psi}^-$) pair is produced just below threshold in the charged Drell-Yan (DY) process and forms a vector bound state $\Upsilon^\pm$ due to the hidden force. The bound state then undergoes annihilation decay into $W^\pm X$ on prompt collider timescales. The motivation for focusing on the electrically charged bound state is twofold: First, its production mediated by $W^\ast$ exchange has larger cross section compared to the neutral channel via $\gamma^\ast / Z^\ast$, and second, selecting the $W\to \ell \nu$ decay provides a hard prompt lepton with sizable branching ratio, ensuring efficient triggering and powerful suppression of the SM backgrounds. 

We assume that $\phi$ decays back to the SM via a small mass mixing with the $125\;\mathrm{GeV}$ Higgs boson, whereas $\gamma_d$ decays via kinetic mixing with the SM photon. We concentrate on the mass region $10\;\mathrm{GeV} \lesssim m_X \lesssim 100\;\mathrm{GeV}$, which offers the best opportunities for detection of the hidden force carriers at the LHC and is motivated by concrete models, for example, of neutral naturalness. Therefore $\phi \to b\bar{b}$ and $\gamma_d \to \ell \ell$ are selected as the most promising final states. We allow for these decays to be either prompt or displaced. 

For prompt $X$ decays, we show that the resonant $\Upsilon^\pm \to W X$ signals can be tested by performing simple extensions of the existing ATLAS and CMS diboson searches. In the case of $(W\to \ell \nu)(\phi \to b\bar{b})$, we show that extending the ATLAS $Wh$ search \cite{Aad:2015yza} to look for $b\bar{b}$ resonances with mass different from $m_h$ provides a powerful coverage. Notice that, in a similar spirit, ATLAS has very recently published a search for resonances that decay into $Xh$, with $X$ a new particle decaying to light quarks \cite{Aaboud:2017ecz}. For $(W\to \ell \nu)(\gamma_d \to \ell \ell)$, where the SM backgrounds are small, we perform a simple estimate based on the ATLAS $WZ$ search \cite{ATLAS-CONF-2013-015,Aad:2014pha} to show the sensitivity to dilepton resonances with mass different from $m_Z$. Our analyses of the $\Upsilon^\pm \to W\phi, \,W\gamma_d$ channels provide further motivation to extend the program of diboson searches to cover resonances that decay into BSM particles. 

For displaced $X$ decays, we propose searches that require a hard prompt lepton from the $W$ in combination with a reconstructed $(b\bar{b})$ or $(\ell \ell)$ displaced vertex. The hard lepton guarantees efficient triggering on the signal events, and the resulting signatures are essentially background-free. We perform simplified projections to estimate the reach achievable at the LHC. 

It is important to emphasize that the resonant \mbox{$\Upsilon^\pm \to W X$} signals studied in this paper allow for the reconstruction of the mass of both the bound state and the hidden force carrier. If we make the assumption that the decay channels available to the bound state are $W X$ and the ``irreducible'' $\bar{f} f^\prime$ (with $f, f^\prime$ SM fermions) mediated by an off-shell $W$, then from the measurement of the signal rate the size of the coupling between the hidden force carrier and the mediators can be inferred. Thus the discovery of the bound state signals would also offer the opportunity to {\it measure} the strength of the hidden force.

After carrying out our collider analyses within the simplified models sketched in Fig.~\ref{fig:feynman}, we apply the results to two explicit, motivated models that contain strongly coupled hidden forces as well as electroweak-charged mediators. This serves as an illustration of the potential impact of the searches we propose. 

The first model example is $\lambda$-SUSY \cite{Barbieri:2006bg}, where the Higgs quartic coupling can be naturally raised by adding to the superpotential a term $\sim \lambda S H_u H_d$, with $S$ a singlet superfield and large $\lambda \sim O(1)$. If the scalar singlet $s$ is light, it mediates a strong force that can lead to the formation of Higgsino bound states at the LHC, which then decay into $Ws$ with large branching fraction. The singlet decays to SM particles via mixing with the Higgs. In this case we thus identify the mediators with the Higgsinos, $\psi \to \tilde{h}$, and the hidden force carrier with the light singlet scalar, $\phi \to s$. This scenario was first discussed in Ref.~\cite{Tsai:2015ugz}. Here we present a more detailed assessment of the future LHC constraints on the model. 

As a second example we consider non-SUSY ultraviolet (UV) extensions of the Twin Higgs model \cite{Chacko:2005pe}, where new vector-like fermions appear that are charged under both the SM and twin gauge symmetries \cite{Chacko:2005pe,Cheng:2015buv}. Some of these exotic fermions, labeled $\mathcal{K}$, carry SM electroweak and twin color charges, and can have masses in the few hundred GeV range without conflicting with experiment or significantly increasing the fine-tuning in the Higgs mass, as discussed in Ref.~\cite{Cheng:2016uqk}. Once they are pair produced in the charged DY process, the exotic fermions form a vector bound state under the twin color force, which can then annihilate into a $W$ plus twin gluons. In the Fraternal version of the Twin Higgs model (FTH) \cite{Craig:2015pha}, the hadronization of the twin gluons can lead to the production of the lightest glueball, which has $J^{PC} = 0^{++}$ and decays into SM particles by mixing with the Higgs. The glueball decay length strongly depends on its mass, and can be either prompt or macroscopic. In this scenario we thus identify the mediators with the exotic fermions, $\psi \to \mathcal{K}$, and the hidden force carrier with the lightest twin glueball, $\phi \to \hat{G}_{0^{++}}$.

Notice that, in the broad setup we are considering, the (scalar or fermion) neutral mediator $\psi^{0}$ can also be the dark matter candidate. The production and decay of the $\psi^0 \bar{\psi}^0$ bound state then gives an example of dark matter annihilation at colliders that does not leave a missing energy signature \cite{Shepherd:2009sa,An:2015pva,Tsai:2015ugz}.

The remainder of this paper is organized as follows. In Sec.~\ref{sec:simpmodel} we analyze the $\Upsilon^\pm \to W X$ processes in the context of simplified models. We perform projections to estimate the LHC sensitivity in the four final states considered, given by $X = \phi$ or $\gamma_d$, each with prompt or displaced decay. We also discuss the sensitivity to the irreducible $\Upsilon^\pm \to \bar{f}f^\prime$ decays, focusing on the cleanest $\ell \nu$ channel, and compare it with the reach in the $\Upsilon^\pm \to W X$ processes. In Sec.~\ref{sec:lambdasusy} we apply our results to the $\lambda$-SUSY model. We show that for large $\lambda$, the signals arising from the charged Higgsino bound state $\Upsilon^\pm_{\tilde{h}}$ have better reach than the standard monojet and disappearing track searches. In addition, in the typical case of prompt $s\to b\bar{b}$ decays the $\Upsilon^\pm_{\tilde{h}} \to W s$ search has better sensitivity compared to $\Upsilon^\pm_{\tilde{h}} \to \ell \nu$. In Sec.~\ref{sec:THiggs} our results are applied to the UV-extended FTH model. Here we find that, even though the branching fraction of the exotic fermion bound state $\Upsilon^\pm_{\mathcal{K}}$ into $W$+$\,$twin glueball is suppressed to the few percent level, this signal provides an interesting complementarity to $\Upsilon^\pm_{\mathcal{K}} \to \ell \nu$ if the lightest twin glueball decays at a macroscopic distance, giving rise to a $(b\bar{b})$ displaced vertex. Our concluding remarks are given in Sec.~\ref{sec:conclusion}.

\section{Simplified model analysis}\label{sec:simpmodel}
In this section we study the LHC sensitivity to the processes
\begin{eqnarray}\label{eq:simplified1}
pp\; \to \; \Upsilon^{\pm}\to \;W^{\pm}\, X\,,
\end{eqnarray}
where $X = \phi, \gamma_d$ decays as
\begin{eqnarray} \label{eq:decays}
\begin{cases}\phi\to b \bar{b} \;\;\qquad(\text{prompt}\,\,\text{or}\,\,\text{displaced}),\\
\gamma_d\to \ell\ell\,\qquad(\text{prompt}\,\,\text{or}\,\,\text{displaced}),\end{cases}
\end{eqnarray}
with $\ell = e, \mu$. Here $\Upsilon^\pm$ (in the following we often drop the electric charge and write just $\Upsilon$) is a bound state with $J^{PC} = 1^{--}$ that carries unit charge under the SM $U(1)_{\text{em}}$, whereas $\phi\,(\gamma_d)$ is a real scalar (real vector) hidden force carrier. As discussed in the Introduction, we make the assumption that $\phi\,(\gamma_d)$ couples to SM particles dominantly through mass mixing with the SM Higgs (kinetic mixing with the SM photon). Then, in the mass region \mbox{$10\;\mathrm{GeV} \lesssim m_{X} \lesssim 100\;\mathrm{GeV}$} the most promising decays of the force carriers are those in Eq.~\eqref{eq:decays}. We study the four types of signals in Eqs.~(\ref{eq:simplified1}) and (\ref{eq:decays}) at the $13$ TeV LHC and set model-independent bounds on $\sigma(\Upsilon)\, \mathrm{BR}(\Upsilon \to W X) \,\mathrm{BR}(X \to F)$, where \mbox{$F = b\bar{b},\, \ell \ell $}, as functions of the masses of the bound state and of the force carrier. We also compare the reach in these channels to that in
\begin{equation}
pp \; \to \; \Upsilon^{\pm}\to \; \ell\, \nu\,,
\end{equation}
which constitutes the irreducible signal of spin-$1$ electroweak-charged bound states. 

Since in Secs.~\ref{sec:lambdasusy} and \ref{sec:THiggs} we interpret our results in explicit models, it is useful to summarize the formulas that give the $\Upsilon^\pm$ production cross section and branching ratios as functions of the underlying parameters. Given two Dirac fermions $\psi^{-, 0}$ with approximately degenerate mass $m_\psi$ and coupled to the SM $W$ boson as $(g/\sqrt{2}) v_\psi^W \bar{\psi}^+ \slashed{W}^-  \psi^0 + \mathrm{h.c.}$, the cross section for production of their vector bound state $\Upsilon^+$ in quark-antiquark annihilation is
\begin{equation}\label{eq:production}
\sigma_{u\bar{d}\to\Upsilon^{+}}=\pi^3\frac{|\psi(0)|^2}{3 m_{\psi}^3} N_c^\prime  \left(\frac{\alpha_W v_{\psi}^W}{1-\frac{m_W^2}{4m_{\psi}^2}}\right)^2 \frac{1}{s} L_{u\bar{d}}\left(\frac{4m_{\psi}^2}{s}\right),
\end{equation}
where $\alpha_W \equiv g^2/(4\pi)$, $L_{u\bar{d}} (\tau) = \int_\tau^1 (dx/x) [u(x) \bar{d}(\tau/x) + u (\tau/x) \bar{d}(x)]$ is the parton luminosity, $s$ is the collider center of mass energy, and the bound state mass was approximated with \mbox{$M_\Upsilon \simeq 2\,m_\psi$}. An analogous expression holds for the production of the charge conjugate $\Upsilon^{-}\,$. The factor $N_c^\prime$ in Eq.~\eqref{eq:production} accounts for the number of hidden degrees of freedom: for example, $N_c^\prime = 1$ if $\psi^{-, 0}$ are identified with the Higgsinos, while $N_c^\prime = 3$ in the case of exotic fermions that transform in the fundamental of a confining hidden $SU(3)$. For definiteness, henceforth we assume $v_\psi^W = 1\,$, which applies for both the Higgsino and exotic fermion bound states. $\psi(0)$ is the wavefunction at the origin, whose value depends on the details of the hidden force. In the Coulomb approximation we have 
\begin{equation} \label{eq:wavefunction}
\frac{|\psi(0)|^2}{m_{\psi}^3} = \frac{C^3 \alpha_{\lambda}^3}{8\pi}\,, 
\end{equation}
where $\alpha_{\lambda} \equiv \lambda^2 / (4\pi)$ is the hidden force coupling strength, and $C$ is a model-dependent constant. For an $SU(N)$ hidden force, \mbox{$C = C_{\psi} - C_{\Upsilon}/2$}, where $C_\psi\,(C_\Upsilon)$ is the quadratic Casimir of the representation where $\psi\,(\Upsilon)$ transforms (see e.g. Refs.~\cite{Kats:2009bv,Kats:2012ym}). For a $U(1)$- or scalar-mediated force, we can instead set $C = 1$ provided the charges are absorbed in the definition of the force coupling strength $\alpha_\lambda$. In these cases, if the force carrier is not massless the formation of bound states can happen only if its wavelength is larger than the Bohr radius, namely $1/m_{X} > 2/(\alpha_\lambda m_\psi)$, or equivalently $m_{X} < m_{\psi} \alpha_{\lambda}/2 \simeq M_\Upsilon \alpha_\lambda/4$.

In this paper we consider scenarios with small mass splitting between $\psi^\pm$ and $\psi^0$, \mbox{$0 < \Delta m_{\psi} = m_{\psi^{\pm}}-m_{\psi^0}\ll m_W$}. The bound state annihilation rate is $\Gamma_{\Upsilon} = N_c^\prime C^3 \{ \alpha_{\lambda}^4\alpha_W / 24 ,\, \alpha_{\lambda}^3\alpha_W^2 / 4 \}\, m_{\psi}$ depending on whether the dominant channel is $\Upsilon \to W \phi$ via a coupling $\lambda \phi (\bar{\psi}^+ \psi^- + \bar{\psi}^0 \psi^0)$, as in $\lambda$-SUSY, or $\Upsilon \to W^* \to \bar{f}f'$, as in the UV-extended FTH \footnote{Notice that in the former case we have assumed that $\alpha_\lambda$ does not run below the scale $m_\psi$, as it is the case in $\lambda$-SUSY.}. In order for the bound state annihilation to take place before the charged constituent decays as \mbox{$\psi^{\pm}\to (W^\ast \to \bar{f} f') \psi^0$}, $\Gamma_{\Upsilon}$ must be larger than
\begin{equation}
\Gamma(\psi^{\pm}\to\psi^0 \bar{f} f') \simeq \frac{3G_F^2 (\Delta m_{\psi})^5}{5\pi^3}\,.
\end{equation}
This sets an upper bound on the mass splitting (for $\Delta m_{\psi}\ll m_W$)
\begin{equation}\label{eq:masssep}
\frac{\Delta m_{\psi}}{m_{\psi}} < 0.16\, (N_c^\prime C^3)^{1/5} \left(\frac{\alpha_\lambda}{0.2}\right)^{\{4,3\}/5} \left(\frac{300\,\text{GeV}}{m_{\psi}}\right)^{4/5}.
\end{equation}
In the region $m_{\psi} > 300\;\mathrm{GeV}$ that we consider in this work, the existing disappearing track constraint \cite{ATLAS-CONF-2017-017,CMS:2014gxa} applies if $c \tau_{\psi^{\pm}} > 0.1$ ns, corresponding to mass splittings smaller than those typically found in our parameter space.

\subsection{$\Upsilon^{\pm}\to W^{\pm}\,\phi\;$ with prompt $\phi\to b\bar{b}$}\label{sec:prombb}
In this case the LHC sensitivity can be estimated by adapting the strategy used in the search for resonances that decay into $(W \to \ell \nu) (h\to b\bar{b})$ \cite{Aad:2015yza}, to allow for an invariant mass of the $b\bar{b}$ pair different from $m_h$. 

The signal is simulated using a simple FeynRules \cite{Alloul:2013bka} model of a charged spin-1 resonance coupled to SM quarks as $\bar{u}\gamma^\mu P_L d \Upsilon^+_\mu + \mathrm{h.c.}$ and to $W\phi$ as $\phi W^{- \mu} \Upsilon^+_\mu + \mathrm{h.c.}$. For both the signal and backgrounds, we generate parton level events with MadGraph5~\cite{Alwall:2014hca}, shower them using PYTHIA6~\cite{Sjostrand:2006za} and pass the result to Delphes3~\cite{deFavereau:2013fsa} for the detector simulation. We adopt most of the Delphes3 configurations proposed in the Snowmass 2013 energy frontier studies~\cite{Avetisyan:2013onh,Anderson:2013kxz}. However, since the $b$-tagging performance has recently been improved by employing multivariate techniques~\cite{ATLAS:2014jfa}, in our analysis we assume the $b$-tagging efficiency to be $70\%$, with $1\%$ rate for a light flavor jet to be mis-tagged as $b$-jet. Jets are reconstructed using the anti-$k_T$ algorithm with distance parameter $R=0.5$.

In the event selection we require the (sub-)leading $b$-jet to have $p_T > 100\,(30)$~GeV and $| \eta^b | < 2.5$. To suppress the $t\bar{t}$ background, we also impose that $N_j \leq 3$, where $N_j$ is the number of jets. In addition, the selection requires one lepton with $p_{T}^{\ell} > 30\;\mathrm{GeV}$ and $| \eta^\ell | < 2.5$, as well as $\slashed{E}_T > 30\;\mathrm{GeV}$, where $\slashed{E}_T$ is the modulus of the missing transverse energy (MET) vector. The MET vector is identified with the neutrino transverse momentum, and the reconstructed transverse mass and transverse momentum of the $W$ must satisfy $m_T^W \in [10, 100]\,\mathrm{GeV}$ and $p_T^W > 200\;\mathrm{GeV}$, respectively \footnote{The $W$ transverse mass is defined as $m_T^W = \sqrt{2[\slashed{E}_T p_{T}^{\ell} (1-\cos\Delta\phi)]}$, where $\Delta\phi$ is the azimuthal separation between the MET vector and the lepton momentum.}. To identify the force carrier $\phi$ we require $m_{bb} - m_\phi \in [-15, 10]\,\mathrm{GeV}$. In order to reconstruct the full $4$-momentum of the $W$ candidate, we extract the longitudinal component of the neutrino momentum by solving $(p_\nu+p_\ell)^2=m_W^2$ \footnote{Following Ref.~\cite{Aad:2015yza}, if the quadratic equation has two real solutions for $p^{z}_\nu$, then we take the one with smaller absolute value. If the solutions are complex, we take the real part.}. This allows us to calculate the invariant mass of the $W bb$ system for each event. In addition, to improve the resolution on $M_{Wbb}$ we apply a standard kinematic fitting procedure that corrects the $b$-jet momenta by imposing $(p_{b_1} + p_{b_2})^2 = m_\phi^2$ (for more details on the procedure, see for example the CMS search for resonances decaying into $hh$ \cite{CMS:2014eda}). 
\begin{figure}
\includegraphics[width=7cm]{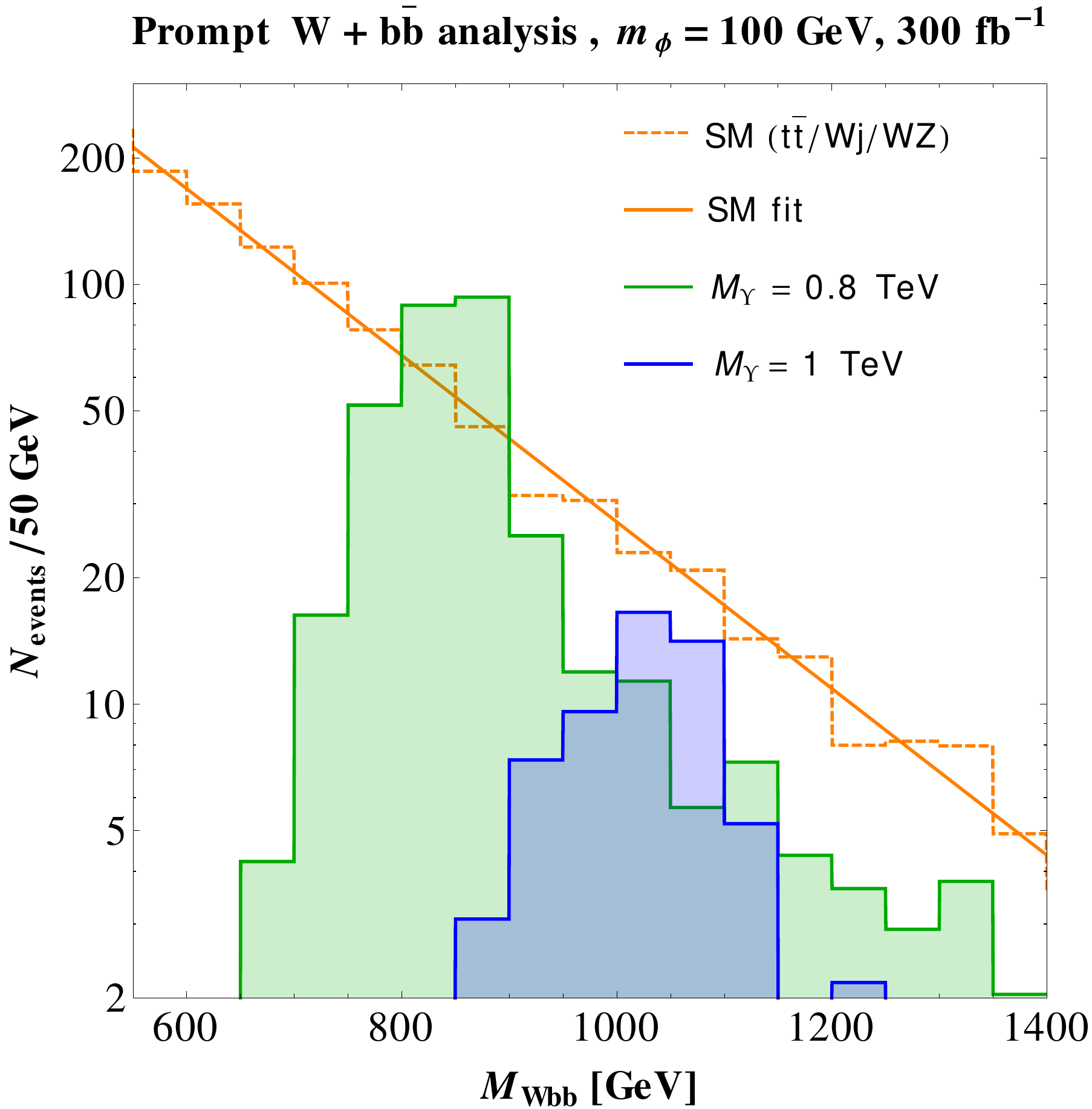}
\caption{Reconstructed $M_{Wbb}$ for signal and backgrounds in the analysis of $\Upsilon^\pm \to W \phi$ with prompt $\phi \to b\bar{b}$, assuming $m_\phi = 100$ GeV. The signal distributions are shown for two representative parameter points with $M_\Upsilon = 800,\,1000$ GeV, taking the hidden force coupling $\lambda=2.5$ and $C = 1$. The orange curve shows the fitted total background that was used to calculate the signal significance.}
\label{figure:lam_wbb}
\end{figure}

The largest SM background is $t\bar{t}$, followed by $W+\,$jets. We also include $W (Z \to b\bar{b})$ production, but its contribution is subdominant. In the calculation of the signal significance, the $M_{W bb}$ distribution of the total background is fitted with an exponential function, shown by the orange curve in Fig.~\ref{figure:lam_wbb}. For the signal, the width of the $M_{Wbb}$ peak is dominated by detector effects and insensitive to the small intrinsic width of the resonance. We then require $M_{Wbb} \in [M_\Upsilon - 50\;\mathrm{GeV},\,M_\Upsilon+100\;\mathrm{GeV}]$.

\begin{figure*}
\center{
\includegraphics[width=7cm]{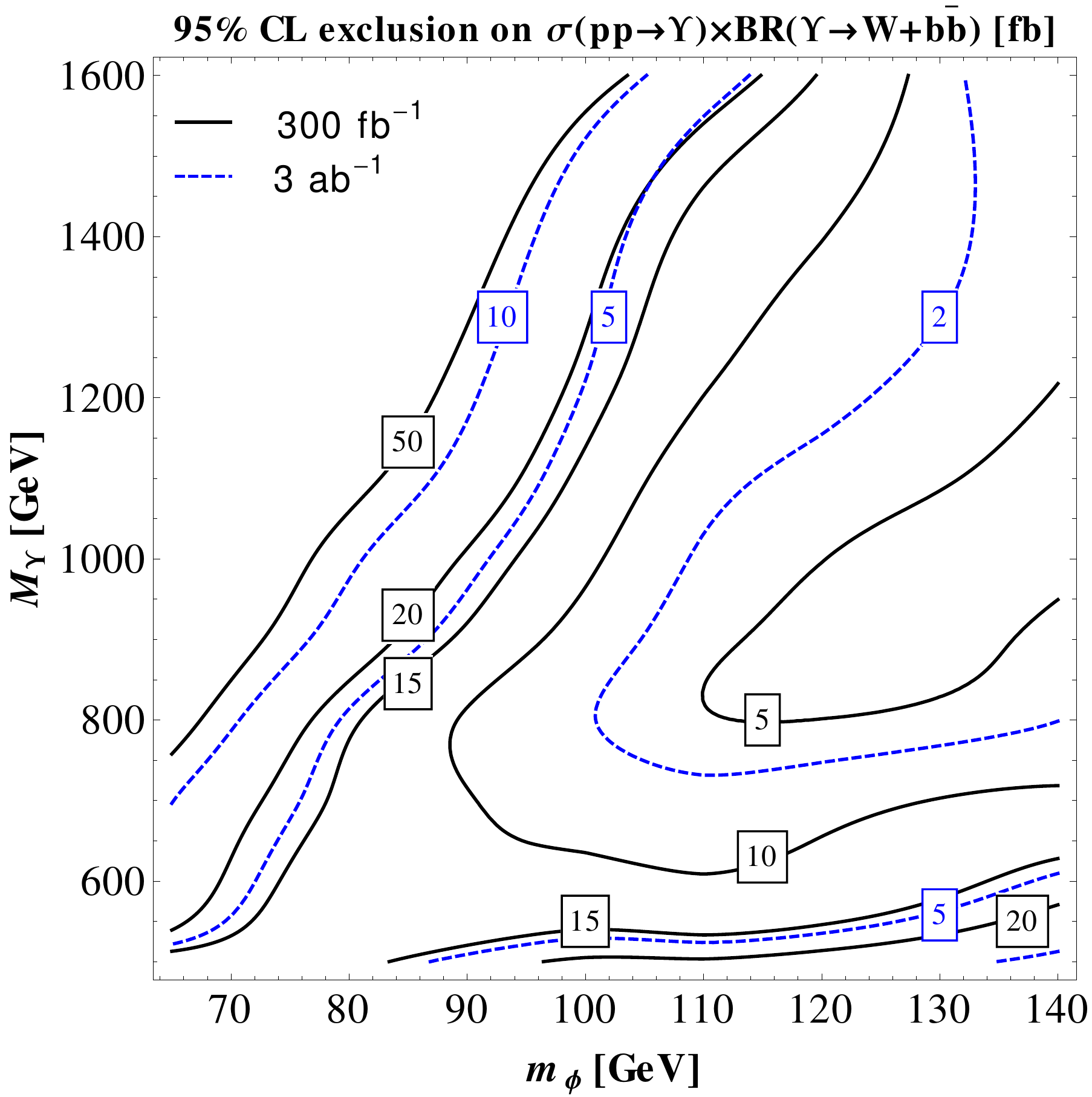}\qquad\quad\includegraphics[width=7cm]{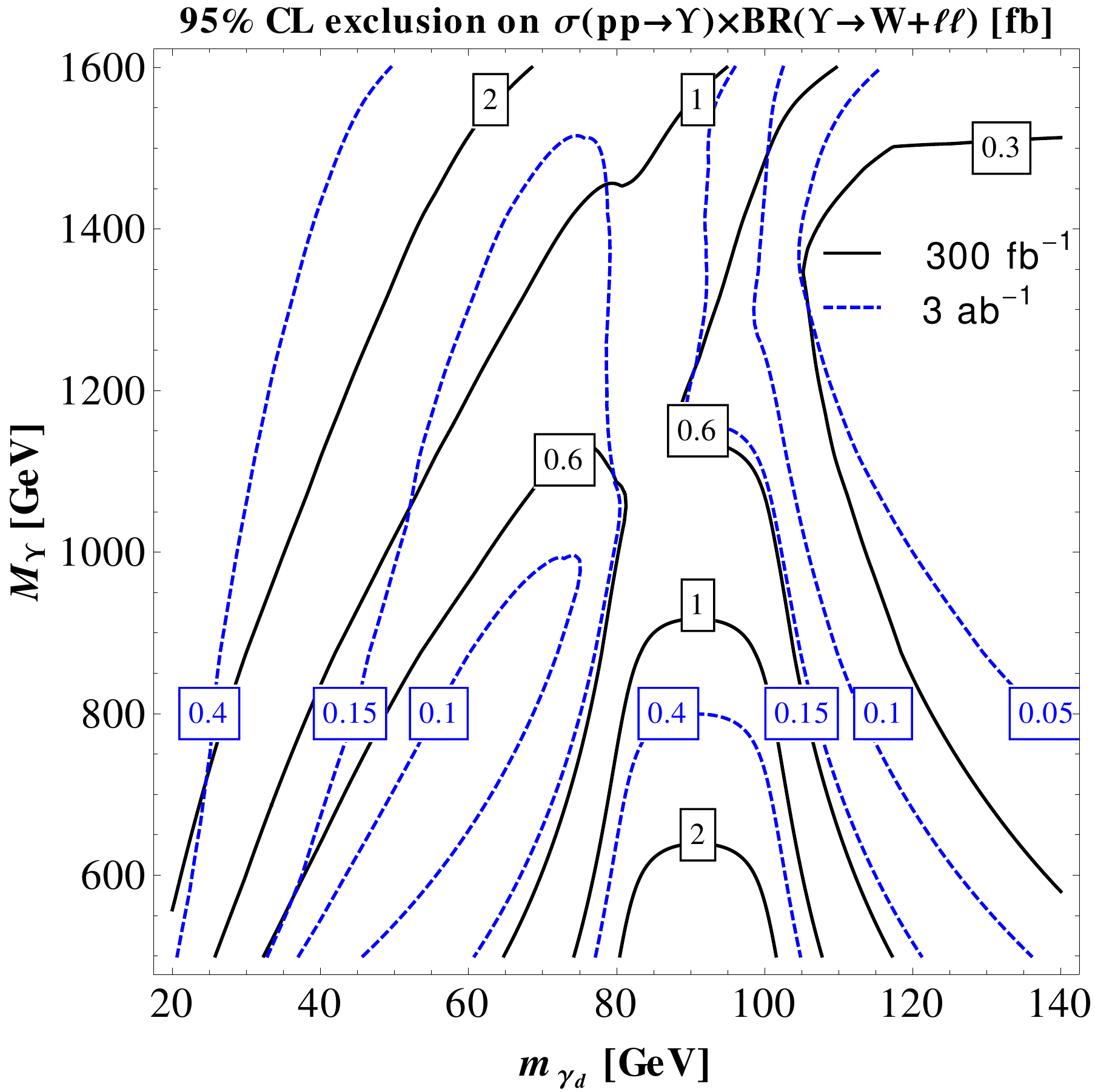}
}
\caption{{\it Left:} Projected $95\%$ CL bounds on $\sigma(\Upsilon) \mathrm{BR}(\Upsilon \to W\phi) \mathrm{BR}(\phi \to b\bar{b})$ from the LHC search for prompt $\phi \to b\bar{b}$ and $W\to \ell \nu$. See Sec.~\ref{sec:prombb} for details. {\it Right:} Projected $95\%$ CL bounds on $\sigma(\Upsilon) \mathrm{BR}(\Upsilon \to W\gamma_d) \mathrm{BR}(\gamma_d\to ee + \mu\mu)$ from the LHC search for prompt $\gamma_d \to \ell \ell$ and $W\to \ell^{\,\prime} \nu$. See Sec.~\ref{sec:trilepton} for details. In both analyses, the invariant mass cuts were varied according to the hypothetical masses of the BSM particles. However, the cross section limits were computed using local, and not global, significance.}
\label{fig:signal}
\end{figure*}

The resulting bounds on $\sigma(\Upsilon)\mathrm{BR}(\Upsilon \to W \phi) \mathrm{BR}(\phi \to b\bar{b})$ are shown as contours in the $(m_\phi, M_\Upsilon)$ plane in the left panel of Fig.~\ref{fig:signal}. We stress that although we have imposed different cuts on the $bb$ and $Wbb$ invariant masses for each hypothetical combination of $(m_\phi, M_{\Upsilon})$ considered, the bounds were calculated using local, and not global, significance. It can be clearly seen that for a fixed $M_\Upsilon \gtrsim 800\;\mathrm{GeV}$, the cross section limit deteriorates when $m_\phi$ is decreased. This happens because in our analysis we require two separate $b$-jets with $\Delta R_{bb}\gtrsim 0.5$, which significantly reduces the selection efficiency for large $M_\Upsilon$ and light $\phi$. For this reason we chose to show our results only for $m_\phi > 60\;\mathrm{GeV}$, below which the efficiency becomes very small \footnote{Notice also that if $m_\phi < m_h/2 = 62.5\;\mathrm{GeV}$, important, albeit model-dependent, constraints can arise from the $h \to \phi \phi$ decay.}. The sensitivity can be extended to larger $M_\Upsilon$ and smaller $m_\phi$ through the application of jet substructure techniques \cite{Butterworth:2008iy}, which go beyond the scope of this paper but can be efficiently implemented in the actual experimental analysis, similarly to the very recent ATLAS searches for resonances decaying to $(W \to q\bar{q}^{\,\prime})(h\to b\bar{b})$ \cite{Aaboud:2017ahz} and $(X \to q\bar{q}^{\,\prime})(h\to b\bar{b})$ \cite{Aaboud:2017ecz}.

\subsection{$\Upsilon^{\pm}\to W^{\pm}\,\gamma_d\;$ with prompt $\gamma_d \to \ell \ell$}\label{sec:trilepton}
The projected bounds on the prompt \mbox{$(\gamma_d \to \ell \ell)(W\to \ell^{\,\prime} \nu)$} signal are obtained by rescaling the results of the $8$ TeV ATLAS search for $WZ$ resonances in the tri-lepton channel \cite{ATLAS-CONF-2013-015,Aad:2014pha}. Notice that even though one neutrino is present in the final state, the kinematics can be fully reconstructed \cite{Aad:2014pha} by solving the equation $(p_\nu + p_{\ell^{\prime}})^2 = m_W^2$ for $p_\nu^z$, with the same procedure described in Sec.~\ref{sec:prombb}. 

We summarize here the rescaling procedure. The SM background, which is dominated by $W\,$+$\,Z^{(\ast)}/\gamma^\ast$ production, is very suppressed if the invariant mass of the $\ell \ell$ pair is away from the $Z$ peak. To estimate it we perform a simulation of SM $pp\to W \ell \ell$ at parton level, in the SM at $13$ TeV, and use it to compute for each $m_{\gamma_d}$ hypothesis the ratio $r(m_{\gamma_d})$ of the cross section in an invariant mass window $| m_{\ell \ell} - m_{\gamma_d} | < 10\,\mathrm{GeV}$ to the cross section on the $Z$ peak, namely $| m_{\ell \ell} - m_{Z} | < 10\,\mathrm{GeV}$. Then, the total background given as a function of $M_{WZ}$ in Table~2 of the ATLAS note \cite{ATLAS-CONF-2013-015} is rescaled to a collider energy of $13\;\mathrm{TeV}$ using the $q\bar{q}^{\,\prime}$ parton luminosity, as well as to the appropriate integrated luminosity, and multiplied times $r(m_{\gamma_d})$ to obtain our background prediction as a function of $M_{W \gamma_d}$. The total signal acceptance times efficiency for a $W'$ with mass $800\;\mathrm{GeV}$, $\mathcal{A} \times \epsilon\, (800)$, was given in Table~7 of Ref.~\cite{ATLAS-CONF-2013-015}. To take into account the variation of the invariant mass shape, for $M_\Upsilon$ different from $800\;\mathrm{GeV}$ we multiply $\mathcal{A}\times \epsilon\, (800)$ by the ratio of the maximum values of the corresponding signal templates, shown in Fig.~5 of the same reference. The resulting acceptance times efficiency, which was calculated for LHC energy of $8$ TeV, is employed in our $13$ TeV projection. In addition, we include the effect of the lepton isolation cuts as a function of the boost factor of $\gamma_d\,$, by requiring an angular separation $\Delta R_{\ell \ell} > 0.3$. After including this correction, our estimate of the signal acceptance times efficiency for $m_{\gamma_d}= 60$ GeV varies from $\approx 7\%$ at $M_\Upsilon = 800\;\mathrm{GeV}$ to $\approx 0.7\%$ at $M_{\Upsilon} = 1.5\;\mathrm{TeV}$. Our rescaling method relies on the assumption of a bump-hunt-type search in a narrow $M_{W\ell \ell}$ window around the putative $M_\Upsilon\,$, which is a reasonable approach given the good experimental resolution achievable in this final state. At the same time, however, some caveats apply to the extrapolation of the $8$ TeV analysis to $13$ TeV. In particular, we have implicitly assumed that the variation of trigger thresholds and selection cuts on the leptons and missing energy will not significantly affect our results.

The resulting bounds on  $\sigma(\Upsilon) \mathrm{BR}(\Upsilon \to W\gamma_d) \mathrm{BR}(\gamma_d\to ee + \mu\mu)$ are shown in the right panel of Fig.~\ref{fig:signal}. We again emphasize that they were computed using local significance. The sensitivity is weaker for light $\gamma_d$ and heavy $\Upsilon$, where the leptons from the dark photon decay are collimated, and for $m_{\gamma_d} \sim m_Z$, where the background is largest.


\subsection{$\Upsilon^{\pm} \to W^{\pm}\,\phi\,(\gamma_d)$ with displaced $\phi \to b\bar{b} \;(\gamma_d \to \ell \ell)$}\label{sec:dissimple}
If the hidden force carrier has a macroscopic decay length, we can search for the $\Upsilon^\pm$ signal in final states containing a prompt hard lepton stemming from the $W$ and a displaced $\phi\to b\bar{b}\,$ or $\gamma_d\to\ell \ell$ decay.
\begin{figure*}
\includegraphics[width=8.5cm]{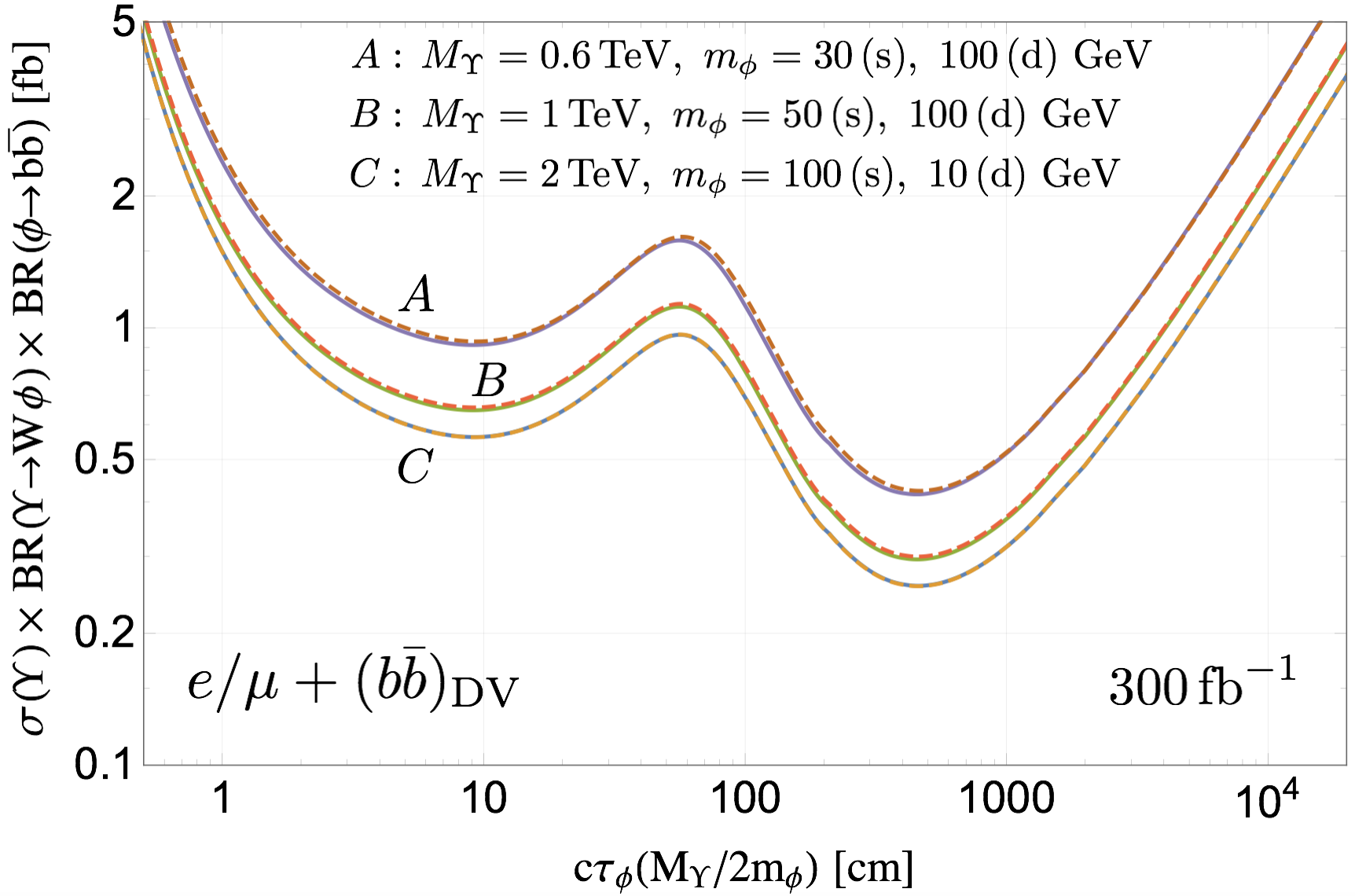}\qquad\includegraphics[width=8.5cm]{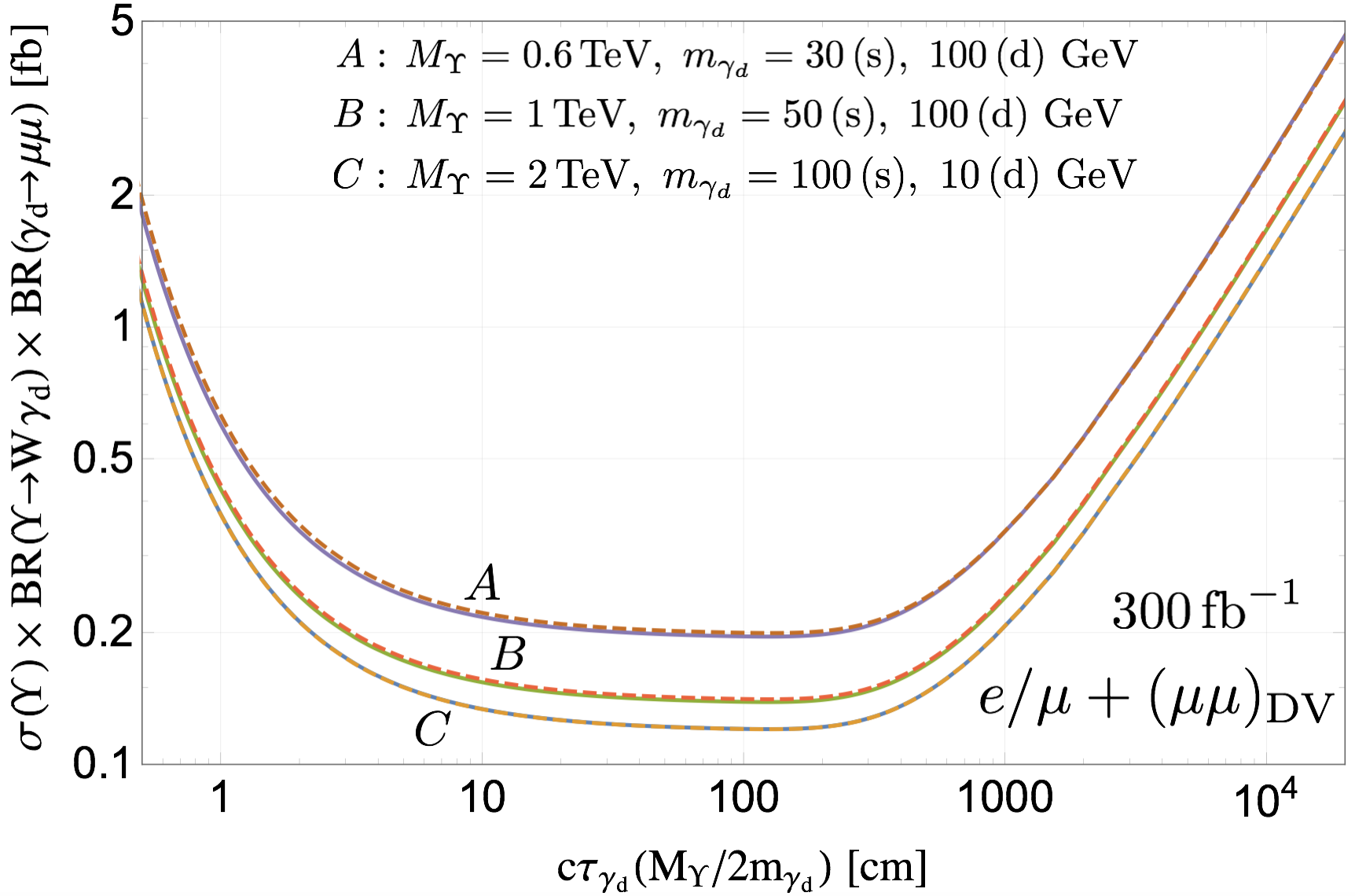}
\caption{{\it Left:} $95\%$ CL upper bound on $\sigma(\Upsilon)\mathrm{BR}(\Upsilon \to W \phi) \mathrm{BR}(\phi \to b\bar{b})$ from the LHC search for prompt $W\to \ell \nu$ and displaced $\phi \to b\bar{b}$. The quantity $c \tau_\phi\, M_\Upsilon / (2m_\phi)$ approximates the decay length of $\phi$ in the lab frame. The result is insensitive to $m_{\phi}$, as can be seen from the small deviation between the solid~(s) and dashed~(d) curves. The local minimum on the left corresponds to decays inside the ID, while the minimum on the right corresponds to the HCAL+MS. {\it Right:} $95\%$ CL upper bound on $\sigma(\Upsilon)\mathrm{BR}(\Upsilon \to W \gamma_d) \mathrm{BR}(\gamma_d \to \mu \mu)$ from the LHC search for prompt $W\to \ell \nu$ and displaced $\gamma_d \to \mu\mu$. Both analyses are described in Sec.~\ref{sec:dissimple}. We assume the searches to be background-free, hence the cross section bounds for $3000$ fb$^{-1}$ are simply obtained by dividing those in the plots by a factor $10$.}\label{fig:DV}
\end{figure*}

For $\phi \to b\bar{b}$, our analysis follows the discussion in Ref.~\cite{Cheng:2015buv}, which in turn was based on the existing ATLAS searches for hadronic displaced vertices (DV) \cite{Aad:2015asa,Aad:2015uaa}. We generate the signal process at the parton level, and require one prompt lepton $\ell = e, \mu$ with $p_T^\ell > 100\;\mathrm{GeV}$ and $| \eta^\ell | < 2.5$, thus ensuring that the signal events can be easily triggered on. An additional $90\%$ efficiency is assumed for the reconstruction of the prompt lepton. In addition, we require two $b$'s with $| \eta^b | < 2.0$ and $p_T^b > 30\;\mathrm{GeV}$. For each event, we calculate the $4$-momentum of $\phi$ in the lab frame, which together with the proper lifetime $c\tau_\phi$ determines the probability distribution for the location of the displaced decay. The DV can be detected either in the inner detector (ID), if its radial distance $r$ satisfies $1\;\mathrm{cm} < r < 28\;\mathrm{cm}$, or in the hadronic calorimeter (HCAL) and muon spectrometer (MS) if $\,200\;\mathrm{cm} < r < 750\;\mathrm{cm}$. For the efficiency of the DV reconstruction we assume a constant $10\%$ in the ID volume and $40\%$ in the HCAL$+$MS, which are simple approximations of the results given in Refs.~\cite{Aad:2015asa,Aad:2015uaa} \footnote{In the analysis of the twin bottomonium signals of Ref.~\cite{Cheng:2015buv}, the efficiency was very conservatively assumed to be $10\%$ also in the HCAL$+$MS. Based on the results of Ref.~\cite{Aad:2015uaa}, we believe $40\%$ to be closer to the actual experimental performance.}.

Notice that the DVs can be identified even when the angular separation between the $b$-jets is small. In the ID the impact parameter $d_0$ of charged tracks can be exploited, as done in Ref.~\cite{Aad:2015uaa}. We can roughly estimate that for a distance $\sim 10$ cm between the location of the displaced decay and the primary vertex, the requirement $d_0 > 1\;\mathrm{cm}$ \cite{Aad:2015uaa} yields sensitivity to $\phi$'s with boost factor as large as $10$. If the decay is inside the HCAL, the ratio of the energy deposits in the electromagnetic calorimeter and HCAL can be used to identify the signal. A detailed understanding of the dependence of the reconstruction efficiency on the boost factor requires further studies, which are beyond the scope of this paper. Here we simply give an estimate, by assuming the above-mentioned boost-independent values for the efficiency.

The analysis of displaced $\gamma_d \to \ell \ell$ is performed along similar lines. The same cuts and efficiency are applied on the prompt lepton originating from the $W$. We focus on $\gamma_d \to \mu \mu$ decays, requiring the two muons to satisfy $| \eta^{\mu} | < 2.0$ and $p_T^{\mu}>30\;\mathrm{GeV}$. Approximating the results of the searches in Refs.~\cite{CMS:2014hka, ATLAS-CONF-2016-042}, we assume that dimuon DVs can be reconstructed for $1\;\mathrm{cm} < r < 750 \;\mathrm{cm}$ with $40\%$ efficiency.

Although in general the searches for DVs at the LHC suffer from several backgrounds, such as the misidentification of prompt objects and the accidental crossing of uncorrelated tracks, these are strongly suppressed by the additional requirement of a prompt hard lepton. Therefore, in both our DV analyses we assume the background to be negligible, and accordingly we exclude at $95\%$ CL all parameter points that would yield a number of signal events larger than $3$.

Even though each of the signals depends on three parameters, namely the masses $M_\Upsilon$ and $m_{X}$ and the proper decay length $c\tau_{X}$, the problem can be simplified by observing that experimentally, the most important variable is the decay length of the long-lived particle in the lab frame. In the approximation that the $\Upsilon^\pm$ is produced at rest, this is simply given by $c\tau_{X} M_\Upsilon/(2m_{X})$. Figure~\ref{fig:DV}, where the bounds on the signal cross section are shown as functions of $c \tau_{X} M_\Upsilon/(2m_{X})$, confirms that this quantity determines the experimental efficiency to a good accuracy. A subleading dependence on $M_\Upsilon$ can be observed, originating from the cuts on the prompt lepton, whereas varying $m_X$ leaves the efficiency essentially unaffected.

\subsection{$\Upsilon^\pm \to \ell\nu$}
The $\Upsilon^\pm$ has an irreducible decay width into SM fermions, via an off-shell $W$ boson. The most powerful probe of these decays is the $\Upsilon^\pm\to \ell \nu$ channel, where the current upper limit on $\sigma(\Upsilon^\pm) \mathrm{BR}(\Upsilon^\pm \to \ell \nu)$ is of $O(\mathrm{few})$ fb for $M_\Upsilon \sim 1\;\mathrm{TeV}$, based on $36.1$ fb$^{-1}$ \cite{Aaboud:2017efa}. We obtain projections to larger integrated luminosity $L$ by rescaling the current cross section constraint $\propto 1/\sqrt{L}\,$. Even though this procedure is strictly correct only when systematic uncertainties are negligible, we have checked that applying it to the constraint from a previous ATLAS analysis based on $13.3$ fb$^{-1}$ \cite{ATLAS:2016ecs} gives good agreement with the $36.1$ fb$^{-1}$ bound of Ref.~\cite{Aaboud:2017efa}. This justifies our simplified treatment.

\begin{figure}[h!]
\includegraphics[width=7cm]{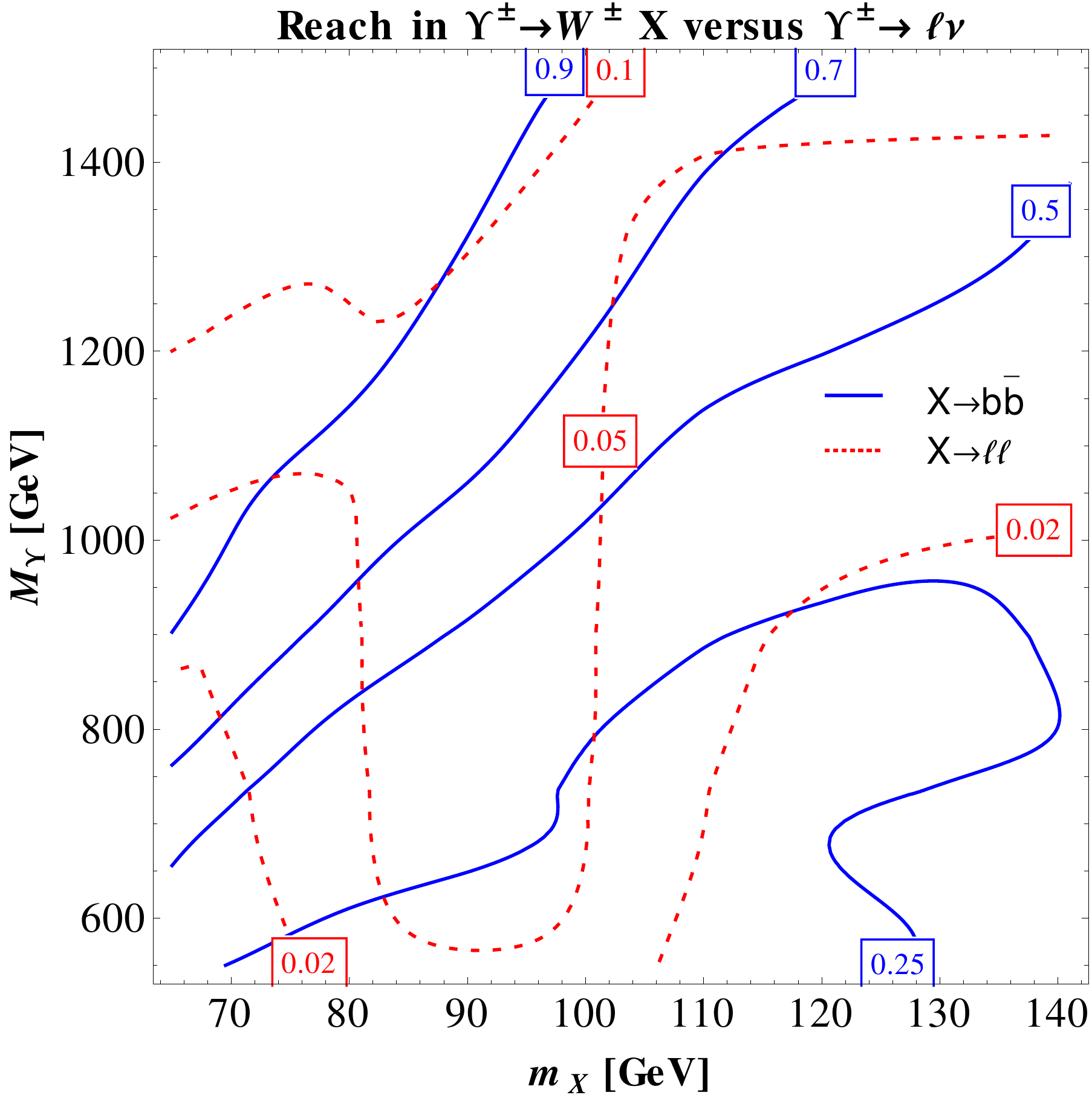}
\caption{Contours of the ratio in Eq.~\eqref{eq:BRratio} that gives at the LHC with $300$ fb$^{-1}$ the same bound on $\sigma(\Upsilon)$ from the $\Upsilon \to WX$ and $\Upsilon \to \ell \nu$ final states. Here we consider prompt $\phi \to b\bar{b}$ and $\gamma_d \to \ell \ell$ decays.}\label{fig:WP}
\end{figure}

It is interesting to compare the sensitivity in the $\Upsilon^\pm \to \ell \nu$ and $\Upsilon^\pm \to W^\pm X$ final states. Focusing on prompt $\phi \to b\bar{b}$ and $\gamma_d \to \ell\ell$ decays, in Fig.~\ref{fig:WP} we show in the $(m_X, M_\Upsilon)$ plane contours of the ratio
\begin{equation} \label{eq:BRratio}
\frac{\mathrm{BR} (\Upsilon^\pm \to W^\pm X)\, \mathrm{BR}(X \to F)}{\mathrm{BR} (\Upsilon^\pm \to W^\pm X)\, \mathrm{BR}(X \to F) + \mathrm{BR} (\Upsilon^\pm \to \bar{f} f^\prime)}
\end{equation}
(where for $\Upsilon^\pm \to \bar{f} f^\prime$ we sum over all SM fermions) that yields with $L= 300$ fb$^{-1}$ the same constraint on $\sigma(\Upsilon^\pm)$ from the $WX$ and $\ell\nu$ final states. For the scalar $\phi$ we find that the ratio in Eq.~\eqref{eq:BRratio} is $< 0.5$ in a large region of parameter space, thus indicating that the search for $\Upsilon \to W \phi$ provides an important test of the bound state properties. On the other hand, the $\Upsilon \to W \gamma_d$ decay can compete with $\Upsilon\to \ell \nu$ even if the relative branching fraction is at the percent level, thanks to the striking trilepton signature.
%

\section{$\lambda\,$-$\,$SUSY}\label{sec:lambdasusy}

Here we discuss the concrete example of $\lambda$-SUSY \cite{Barbieri:2006bg}, where a coupling of the form $\sim \lambda S H_u H_d$ is added to the minimal supersymmetric SM superpotential, with $S$ a singlet scalar superfield. A large $\lambda \sim O(1)$ helps to increase the Higgs mass to $125\;\mathrm{GeV}$ in a natural way \cite{Hall:2011aa}. If in addition the singlet scalar $s$ is light, it mediates a strong attractive force between the Higgsinos, that can lead to the formation of bound states in the process of DY Higgsino pair production \cite{Tsai:2015ugz}. The charged bound state $\Upsilon^\pm_{\tilde{h}}$ decays into $Ws$ with large branching fraction, and in turn the $s$ decays to $b\bar{b}$ through its mixing with the Higgs.  
\begin{figure*}
\includegraphics[trim=0cm -0.47cm 0cm 0cm,width=7.1cm]{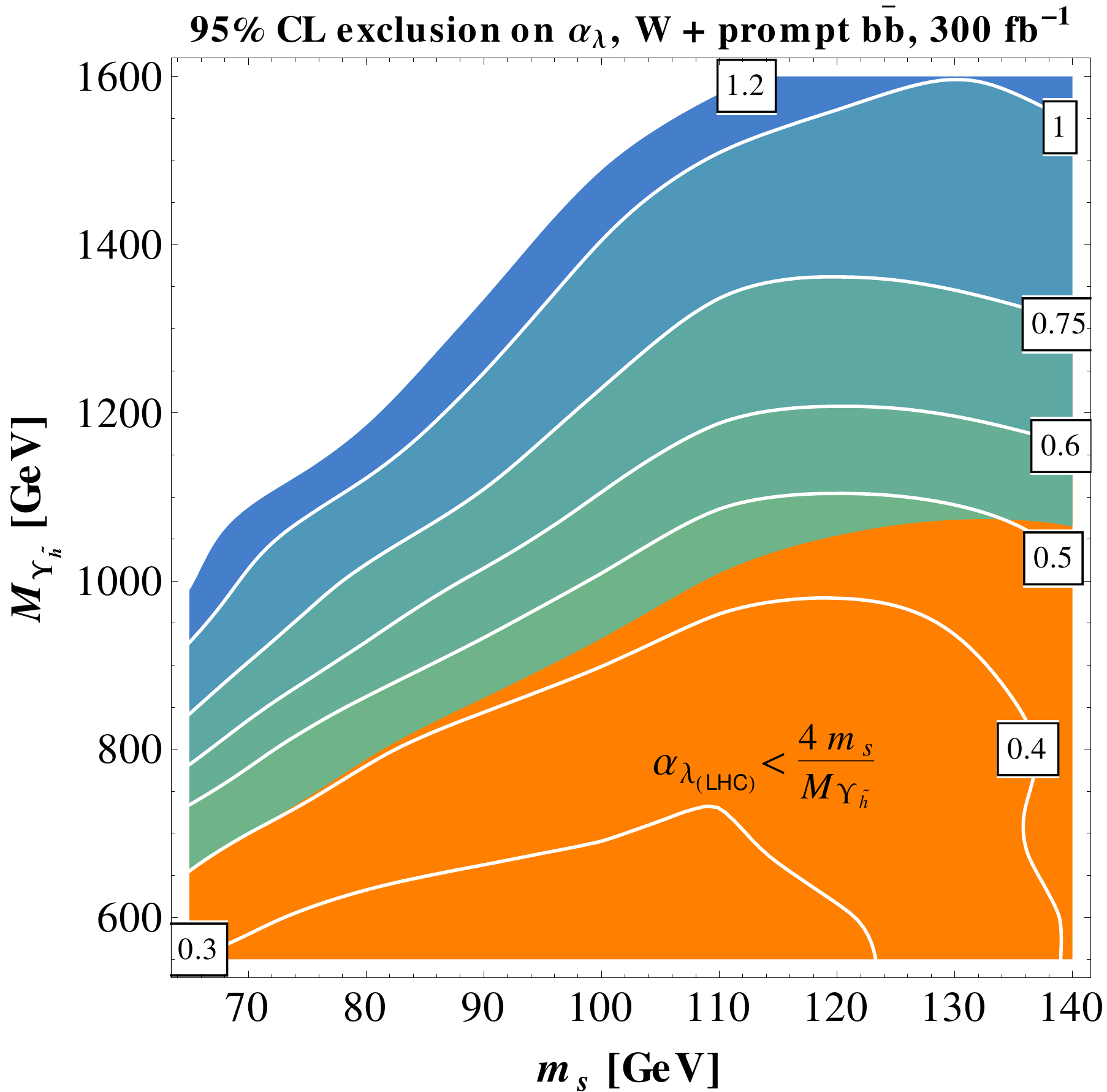} \qquad \quad \includegraphics[trim=0cm -0.3cm 0cm 0cm,width=7.1cm]{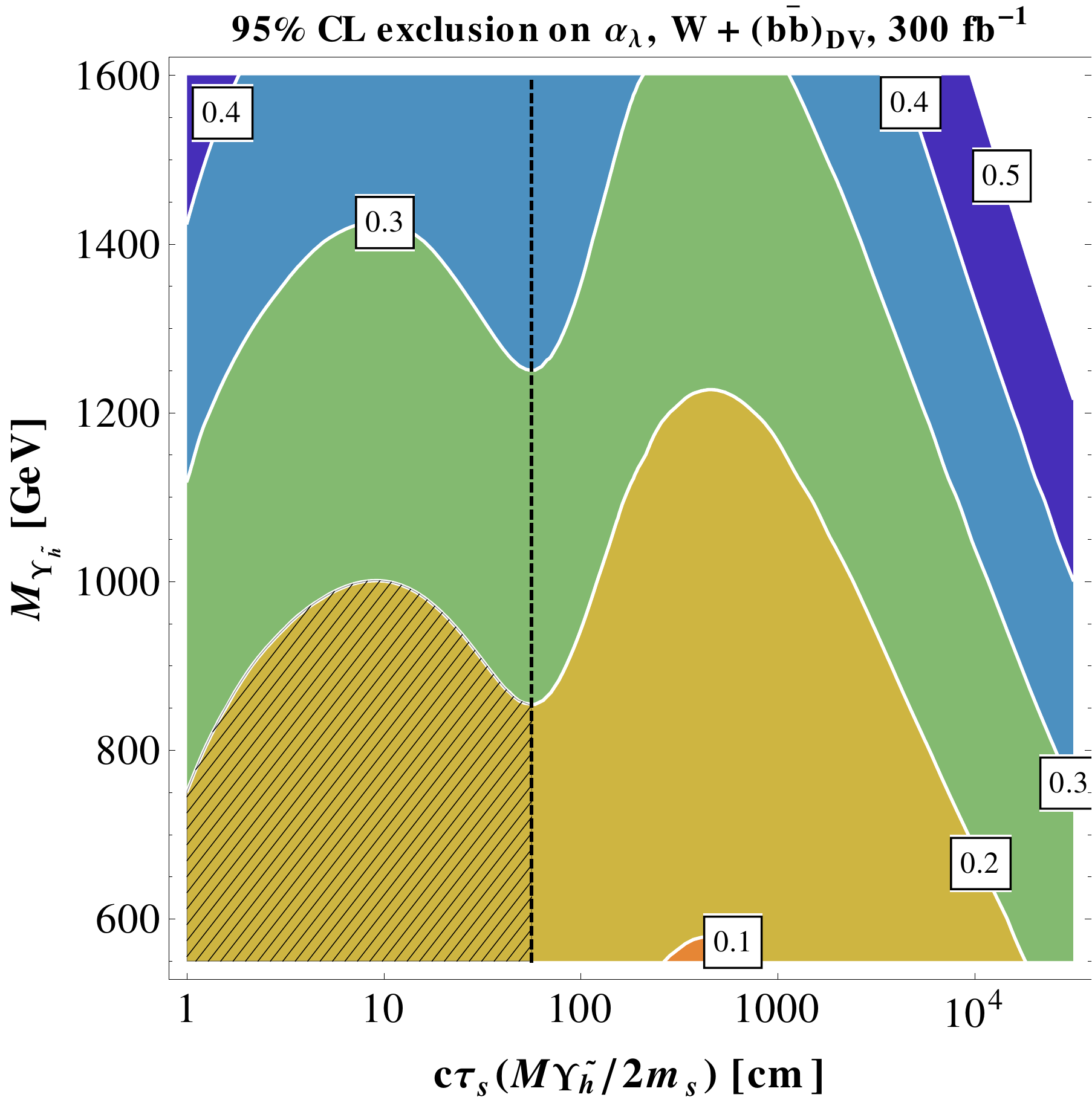}
\caption{{\it Left:} $95\%$ CL upper bound on $\alpha_\lambda$ in $\lambda$-SUSY from the search for prompt $(W\to \ell \nu)(s \to b\bar{b})$ at the LHC with $300$ fb$^{-1}$. In the orange-shaded region the LHC can fully rule out the existence of Higgsino bound states, by setting a limit on $\alpha_\lambda$ that is below the smallest value required for bound state formation, $4 m_{s}/M_{\Upsilon_{\tilde{h}}}$. {\it Right:} $95\%$ CL upper bound on $\alpha_\lambda$ in $\lambda$-SUSY from the search for prompt $W\to \ell \nu$ and displaced $s \to b\bar{b}$ at the LHC with $300$ fb$^{-1}$. The relation $\alpha_\lambda > 4 m_{s}/M_{\Upsilon_{\tilde{h}}}$ that makes bound state formation possible is implicitly assumed to hold. In the region to the left (right) of the vertical dashed line, $s$ decays in the ID (HCAL+MS). In the region hatched in black the $s$ decays in the ID with boost factor $\gtrsim 10$, making the identification of the DV challenging (see text for details).}
\label{fig:Apsearch}
\end{figure*}

Before applying the results of our analysis of Sec.~\ref{sec:simpmodel}, we briefly summarize some essential aspects of the model. We consider a general next-to-minimal supersymmetric SM superpotential \mbox{$W = \sqrt{2} \lambda S H_u H_d + \xi_F S+ \mu' S^2/2+ \kappa S^3/3$} and assume the gauginos to be heavy and out of the LHC reach \footnote{The first term in the superpotential is normalized such that the coupling of the physical scalar singlet $s$ to the Higgsinos is simply $\lambda$.}. We focus on the limit $2 \kappa \langle s \rangle + \mu' \gg \lambda v_{u,d}$ (where we have expanded the scalar component of the superfield $S$ as $s\to \langle s \rangle + s/\sqrt{2}$), so the singlino is also decoupled from the light Higgsinos. As a consequence, the up- and down-type Higgsinos are nearly degenerate, and their DY production is unsuppressed. We can then treat $(\tilde{h}^0_u, \,\tilde{h}^0_d)$ as a Dirac fermion that receives a mass $m_{\tilde{h}}$ from the $\mu$-term, and similarly for the charged Higgsinos. Electroweak radiative corrections split the masses of the neutral and charged Higgsinos by $\Delta m_{\tilde{h}} \simeq 350$ MeV, which clearly satisfies the condition in Eq.~(\ref{eq:masssep}). The singlet scalar $s$ decays into SM particles through its mixing with the SM-like Higgs, which is constrained to be $\lsim 20\%$ by the existing Higgs couplings measurements \cite{Farina:2013fsa}. Since $\lambda$ also generates a large coupling between the Higgs and two singlet scalars, we avoid bounds from the $h\to ss$ decay by requiring $m_s > m_h/2$. Therefore, in our study we focus on the singlet scalar mass range
\begin{equation}
\frac{m_h}{2} < m_s < \frac{\mht \alam}{2}\,,
\label{eqn:lamlimit}
\end{equation}
where the second inequality ensures that the bound state can form, as discussed below Eq.~\eqref{eq:wavefunction}. The decay $h\to s s^* \to 4b$ can easily have a small branching ratio, being suppressed by the $h$-$s$ mixing and by the bottom Yukawa coupling. 

The production and decay of $\Upsilon^\pm_{\tilde{h}}$ is described by the upper diagram in Fig.~\ref{fig:feynman}, with the identifications \mbox{$(\psi^0,\psi^{\pm},\phi) \to (\tilde{h}_{u,d}^0,\tilde{h}_{u,d}^{\pm},s)$}. The $s\to b\bar{b}$ decay is generically prompt, but it can also happen at a macroscopic distance if cancellations between the soft SUSY masses suppress the mixing between $h$ and $s$ to less than $\sim 10^{-5}$. We can then reinterpret our simplified model results in the $\lambda$-SUSY context, by comparing the model-independent limits calculated in Secs.~\ref{sec:prombb} and \ref{sec:dissimple} for the $(W\to \ell \nu)(\phi \to b\bar{b})$ final state (with prompt or displaced $\phi$ decay, respectively) to the production cross section of $\Upsilon_{\tilde{h}}$ calculated via Eqs.~\eqref{eq:production} and \eqref{eq:wavefunction}. We appropriately set $N_c^\prime = 1$ and $C = 1$ in those equations. Since the $\Upsilon^{\pm}_{\tilde{h}}$ can annihilate into both $Ws$ and $\bar{f}f'$, in our signal predictions we include the corresponding branching ratio \mbox{$\mathrm{BR}(\Upsilon_{\tilde{h}} \to W s) \simeq \alpha_\lambda / (\alpha_\lambda + 6 \, \alpha_W)$}. Furthermore, we include the $\mathrm{BR}(s \to b\bar{b})$, which is the same as for a SM Higgs with mass given by $m_s$, because $s$ couples to SM fields only via mixing with the Higgs.

In the left panel of Fig.~\ref{fig:Apsearch} we show the constraints on $\alpha_\lambda$ obtained from the prompt $(W\to \ell \nu)(s \to b\bar{b})$ channel with $300$ fb$^{-1}$. Notice that the $\alpha_\lambda$-contours also give at least a rough idea of the \emph{measurement} of the hidden force coupling that can be obtained if an excess is observed. In the orange-shaded region the LHC will be able to entirely rule out the existence of Higgsino bound states, by pushing the exclusion on $\alpha_\lambda$ below the smallest value that allows bound state formation, namely $\alpha_{\lambda}^{\rm min} = 2 m_s / m_{\tilde{h}} =  4m_s / M_{\Upsilon_{\tilde{h}}}\,$. For example, for $\alpha_{\lambda} = 0.4$ the reach extends up to Higgsino masses $m_{\tilde{h}}\sim 500$ GeV. It is interesting to compare this to the reach of the monojet and disappearing track searches. The monojet channel has a $95\%$ CL reach of $m_{\tilde{h}} \simeq 200$ GeV at the LHC with $3$ ab$^{-1}$, and a similar sensitivity is expected in the disappearing track search if the mass splitting generated by electroweak loops, $\Delta m_{\tilde{h}}\simeq 350$ MeV, is assumed \cite{Arkani-Hamed:2015vfh}. Thus we find that if $\lambda$ is large, the reach of the Higgsino bound state signal is far superior. In addition, since the values of $\alpha_{\lambda}$ probed by the analysis correspond to \mbox{$\mathrm{BR}(\Upsilon_{\tilde{h}}\to Ws) \simeq 0.6\,$-$\,0.8$}, after including $\mathrm{BR}(s\to b\bar{b})$ and comparing with Eq.~\eqref{eq:BRratio} and Fig.~\ref{fig:WP} we find that the $\Upsilon_{\tilde{h}} \to (W\to \ell \nu)(s\to b\bar{b})$ final state has better sensitivity than $\Upsilon_{\tilde{h}} \to \ell \nu$ in this region of parameters. 

\begin{figure}
\includegraphics[width=8.5cm]{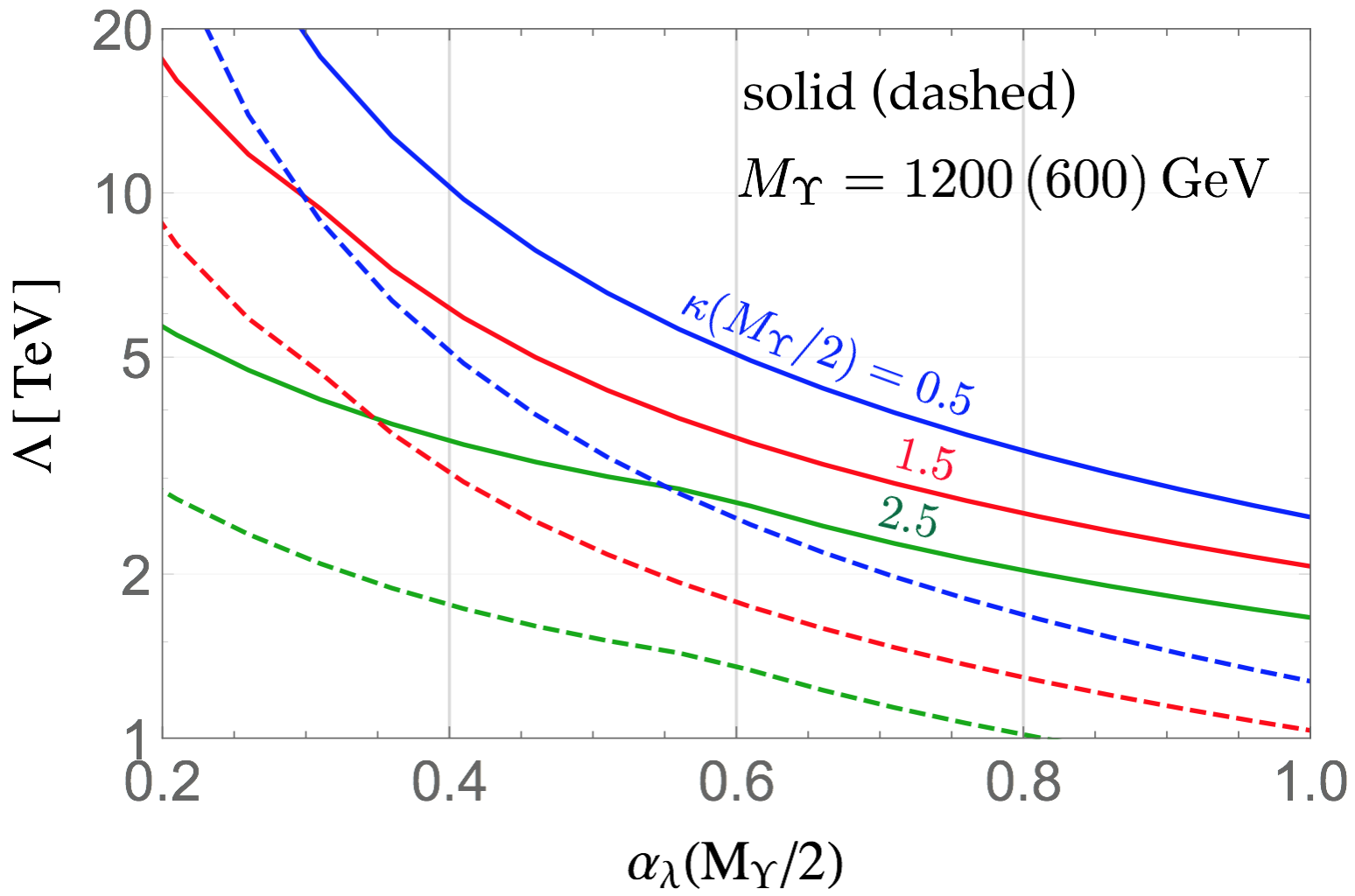}
\caption{Estimate of the scale $\Lambda$ where perturbativity is lost in the $\lambda$-SUSY model, as a function of the low-energy value of $\alpha_\lambda$, for representative values of $\kappa$. $\Lambda$ is defined as the scale where the two-loop contributions to the running of $\lambda$ and $\kappa$ become of the same size as the one-loop terms.}
\label{fig:cutoff}
\end{figure}

For the large values of $\lambda$ that can be probed by our analysis, perturbativity is lost at a relatively low scale $\Lambda$, as illustrated in Fig.~\ref{fig:cutoff}. For example, for $\alpha_\lambda = 0.4$ (corresponding to $\lambda \simeq 2.2$) we find $2\;\mathrm{TeV} \lesssim \Lambda \lesssim 10\;\mathrm{TeV}$, depending on the Higgsino mass and on the value of the parameter $\kappa$ that controls the size of the $S^3$ term in the superpotential. 
The large value of $\lambda$ also affects the Higgs mass prediction. Since the $h$-$s$ mixing is constrained to be small by LHC measurements \cite{Farina:2013fsa}, we have approximately
\begin{equation}
m_{h}^2 \sim \lambda^2v^2\sin^2 2\beta+m_Z^2\cos^2 2\beta\,,
\end{equation}
where $v = \sqrt{ 2(v_u^2 + v_d^2)} \simeq 246$ GeV. Therefore in the region $\lambda \gtrsim 2$ where the bound state production is relevant, $\tan\beta\gtrsim 10$ is required. The $\lambda$-SUSY region with large $\lambda$ and large $\tan\beta$ can produce dangerous corrections to the $S$ and $T$ parameters of electroweak precision tests. Nevertheless, these can be reduced by suppressing the mixing between the Higgsinos and the singlino, as we have assumed from the beginning, and by raising the masses of the squarks and the charged Higgs~\cite{Franceschini:2010qz}.

In the right panel of Fig.~\ref{fig:Apsearch} we show the constraints obtained from the (prompt $W\to \ell \nu)\,$+\mbox{$\,($displaced $s\to b\bar{b}$)} channel. Since the boost factor of $s$ is $\gamma_s \simeq M_{\Upsilon_{\tilde{h}}} / (2m_s)$, the second inequality in Eq.~\eqref{eqn:lamlimit} implies that \mbox{$\gamma_s \gsim 2 \alpha_{\lambda}^{-1}$}. As discussed in Sec.~\ref{sec:dissimple}, the identification of the hadronic DV becomes very challenging if the $s\to b\bar{b}$ decay takes place in the ID with \mbox{$\gamma_{s} \gsim 10$}. This is verified in the region of parameters with $c\tau_s (M_{\Upsilon_{\tilde{h}}} / 2 m_{s})\lsim 30\;\mathrm{cm}$ and $\alpha_{\lambda} \lesssim 0.2$ (hatched in black), where new ideas are required to successfully reconstruct the narrow displaced jet in the ID.

\section{UV-extended Fraternal Twin Higgs}\label{sec:THiggs}
\begin{figure}
\begin{center}
\includegraphics[width=6.5cm]{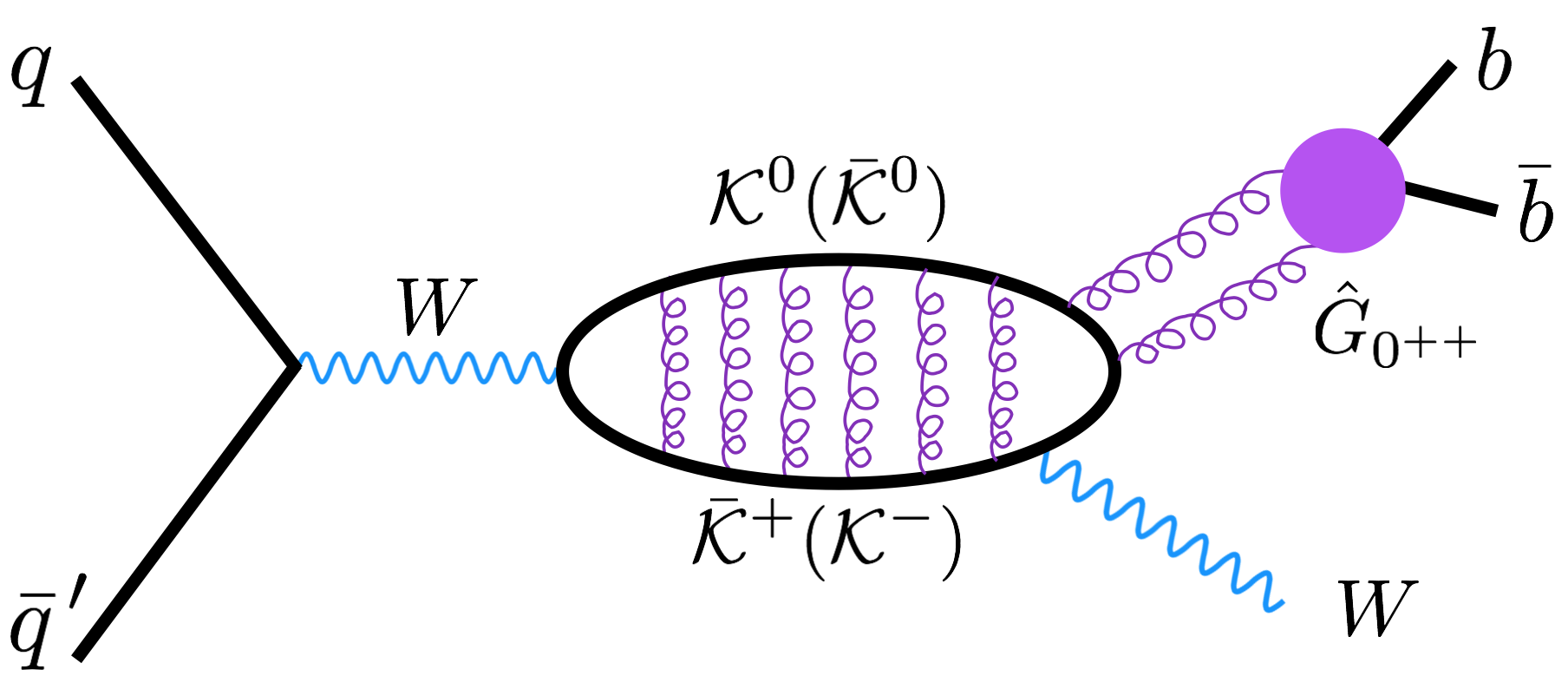}
\end{center}
\caption{The signal of the exotic fermion bound state in the UV-extended Fraternal Twin Higgs. The outgoing twin gluons (curly lines) hadronize into twin glueballs. The lightest glueball $\hat{G}_{0^{++}}$ decays to $b\bar{b}$ through the Higgs portal, either promptly or at a macroscopic distance.}\label{fig:feynTH}
\end{figure}

The signals we study also appear in several non-SUSY UV completions of the TH model, which contain exotic fermions charged under both the SM and twin gauge groups \cite{Chacko:2005pe,Cheng:2015buv}. Some of these fermions, labeled $\mathcal{K}^{-,\,0}$ (where the superscript indicates the SM electric charge), carry SM electroweak and twin color charges. As shown in Ref.~\cite{Cheng:2016uqk}, $\mathcal{K}^{-,\,0}$ can have $Z_2$-breaking masses $\ll 1\;\mathrm{TeV}$ without violating experimental constraints, and without significantly increasing the fine-tuning of the Higgs mass. The exotic fermions can therefore be produced at the LHC through the DY process and form an electrically charged vector bound state $\Upsilon_{\mathcal{K}}^{\pm}$ due to the twin color force. If the lifetime of the constituents is sufficiently long, the bound state annihilates into resonant final states. The main channel is $\bar{f} f'$ via an off-shell $W$, but a sizable branching ratio also exists for the $W \hat{g}\hat{g}$ final state, where the two twin gluons can hadronize into the lightest twin glueball $\hat{G} \equiv \hat{G}_{0^{++}}$; see Fig.~\ref{fig:feynTH}. In turn, the twin glueball decays to $b\bar{b}$ via the Higgs portal, either promptly or at a macroscopic distance depending on the value of the twin confinement scale $\hat{\Lambda}$.

Before we interpret the bounds of Sec.~\ref{sec:simpmodel} in this context, it is useful to recall some important features of the model. The $\mathcal{K}^0$ has a small mass mixing with the twin top. Assuming $m_{\mathcal{K}^{-}} < m_t f / v$, where $f$ is the global symmetry breaking scale, the level repulsion makes $\mathcal{K}^0$ slightly lighter than $\mathcal{K}^-$, with mass splitting \mbox{$m_{\mathcal{K}^-}  = m_{\mathcal{K}^0} + \Delta m_{\mathcal{K}}$} given by 
\begin{equation} \label{eq:exoquarksplit}
\frac{\Delta m_{\mathcal{K}}}{m_{\mathcal{K}^{-}}}\simeq\frac{m_t^2}{2(m_t^2 f^2 / v^2 - m_{\mathcal{K}^{-}}^2)}\,.
\end{equation}
Taking $f/v \simeq 4$ and a typical strength of the twin QCD coupling $\hat{\alpha}_s (q_{\rm rms}) \sim 0.2 $ [where $q_{\rm rms}$ is related to the inverse Bohr radius of the bound state by an $O(1)$ factor \footnote{Precisely, $q_{\rm rms} = (\sqrt{3} \,a_0)^{-1}$, where $a_0 = 2/(C_F \hat{\alpha}_s (q_{\rm rms}) m_{\mathcal{K}^-})$ is the Bohr radius and $C_F = 4/3$ \cite{Kats:2012ym}.}, and we have assumed $\hat{\Lambda} = 5$ GeV], the mass splitting in Eq.~\eqref{eq:exoquarksplit} satisfies Eq.~(\ref{eq:masssep}) when the bound state mass is $M_{\Upsilon_{\mathcal{K}}} < 1.2$ TeV. On the other hand, the neutral exotic fermion $\mathcal{K}^0$ decays into $\hat{W}\hat{b}$, where the twin $W$ can be on- or off-shell, with amplitude suppressed by a mixing angle $\simeq v/f$  (for $m_{\mathcal{K}^-} \ll m_t f/v$). If $m_{\mathcal{K}^0} < m_{\hat{W}} + m_{\hat{b}}$, the corresponding lifetime is sufficiently long to allow for annihilation of the charged bound state. However, the twin bottom cannot be too heavy, to avoid introducing a new source of significant fine-tuning in the Higgs mass. Requiring this additional tuning to be better than $10\%$ restricts the parameter space for the bound state signals to $M_{\Upsilon_{\mathcal{K}}} < 1.1\;\mathrm{TeV}$, which we assume in the following.

Lattice computations~\cite{Chen:2005mg} give $m_{\hat{G}} \simeq 6.8 \hat{\Lambda}$ for the mass of the lightest glueball. The twin confinement scale depends on the number of flavors in the twin sector, as well as on the value of the twin QCD coupling in the UV, $\hat{g}_s (\Lambda_{\rm UV})$, where for definiteness we take \mbox{$\Lambda_{\rm UV} = 5\;\mathrm{TeV}$}. As to the field content, here we focus on the Fraternal Twin Higgs model, which includes twin copies of the third-generation fermions only. Concerning the value of $\hat{g}_s$, assuming exact $Z_2$ symmetry at $\Lambda_{\rm UV}$ leads to \mbox{$\hat{\Lambda} \simeq 5$ GeV}, whereas allowing for a $10\%$ difference between $g_s(\Lambda_{\rm UV})$ and $\hat{g}_s(\Lambda_{\rm UV})$ yields $\hat{\Lambda}\in [1, 20]\;\mathrm{GeV}$, and therefore a lightest glueball mass in the range $7\;\mathrm{GeV} \lesssim m_{\hat{G}} \lesssim 140\;\mathrm{GeV}$. The $\hat{G}$ mixes with the SM-like Higgs $h$ through a twin top loop. In the region of larger mass, $60\;\mathrm{GeV} \lesssim m_{\hat{G}} \lesssim 140\;\mathrm{GeV}$, it decays promptly, hence the dilepton channel $\Upsilon_{\mathcal{K}} \to \ell \nu$ \cite{Cheng:2016uqk} has far better sensitivity than $\Upsilon_{\mathcal{K}} \to W \hat{G}$ due to the much larger branching fraction. Instead, a lighter glueball with mass $15\;\mathrm{GeV} \lesssim m_{\hat{G}} \lesssim 50\;\mathrm{GeV}$ undergoes displaced decays within the volume of the LHC detectors, yielding a signature that is striking enough to potentially overcome the branching fraction suppression. In this mass region the decay is dominantly into $b\bar{b}$, with proper lifetime that can be approximated as \cite{Craig:2015pha}
\begin{equation}
c\tau_{\hat{G}}\sim 1\,\text{cm}\left(\frac{5\,\text{GeV}}{m_{\hat{G}}/6.8}\right)^7\left(\frac{f}{1\,\text{TeV}}\right)^4.
\end{equation}
We then proceed to apply the bound from the \mbox{(prompt $W\to \ell \nu)\,$}+\mbox{$\,($displaced $\phi \to b\bar{b}$)} analysis that was presented in Sec.~\ref{sec:dissimple}, with the identification $\phi \to \hat{G}$.
\begin{figure}
\begin{center}
\includegraphics[width=7cm]{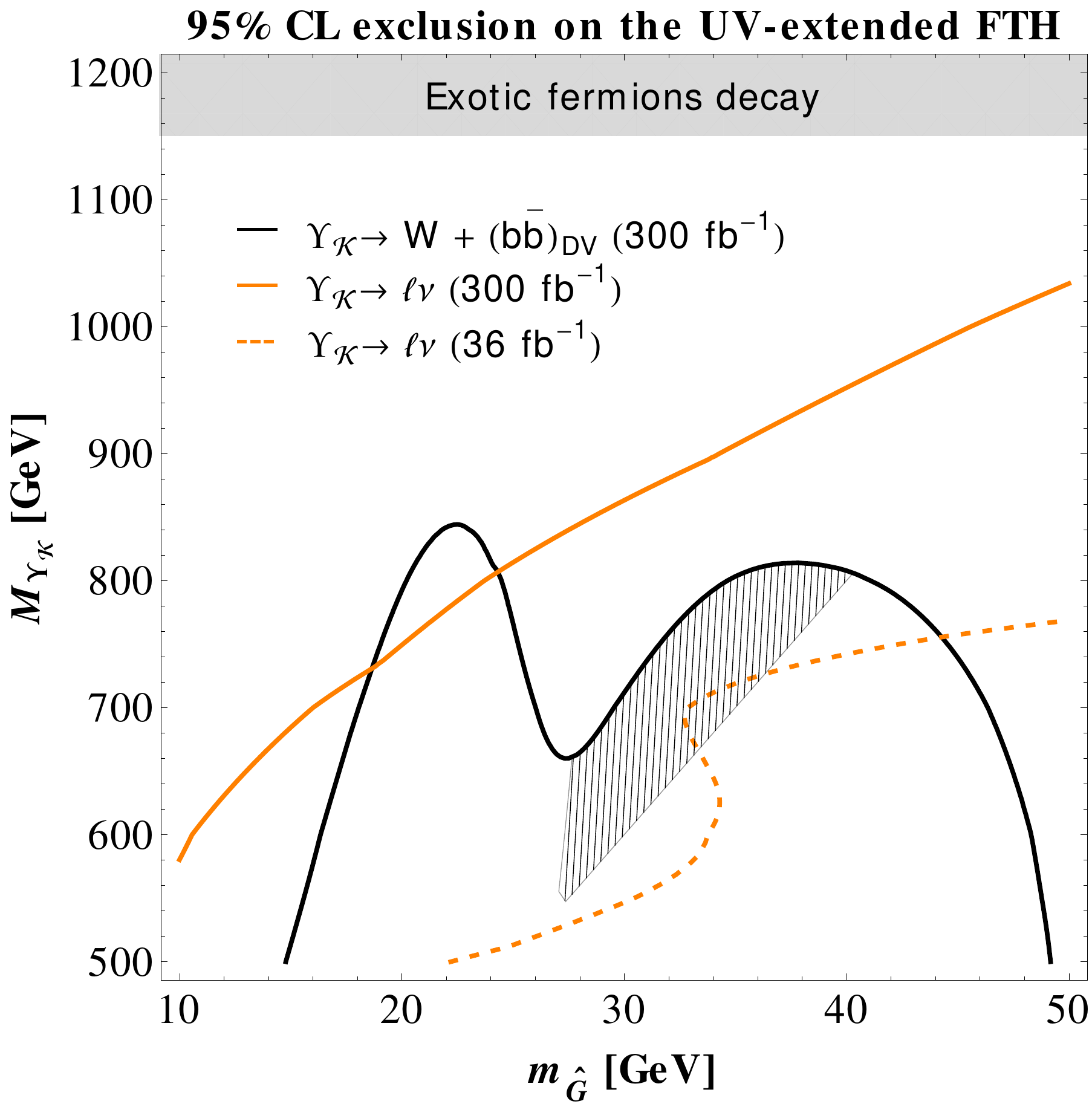}
\end{center}
\caption{$95\%$ CL exclusions on the UV-extended Fraternal Twin Higgs from the LHC searches for signals of the exotic fermion bound state $\Upsilon^\pm_{\mathcal{K}}$. We set $f = 1\;\mathrm{TeV}$. In black, we show the exclusion from the search for prompt $W\to \ell \nu$ and displaced $\hat{G} \to b\bar{b}$ with $300$ fb$^{-1}$. The maximum of reach on the left~(right) corresponds to $c \tau_{\hat{G}}(M_{\Upsilon_{\mathcal{K}}}/ 2 m_{\hat{G}}) \sim 400\,(10)\;\mathrm{cm}$, which optimizes the sensitivity in the HCAL+MS (ID). In the region hatched in black the $\hat{G}$ decays in the ID with boost factor $\gtrsim 10$, making the identification of the DV challenging (see text for details). In orange, we show the current and projected exclusions from the search for $\Upsilon^\pm_{\mathcal{K}} \to \ell \nu$. In the area shaded in grey, the exotic quarks decay before the bound state annihilation takes place.}\label{fig:THbound}
\end{figure}

The cross section for $\Upsilon_{\mathcal{K}}$ production is given by Eqs.~\eqref{eq:production} and \eqref{eq:wavefunction} after we set $N_c^\prime = 3$, $C = 4/3$ and replace $\alpha_\lambda \to \hat{\alpha}_s\,$. To estimate the relative branching ratio for the \mbox{$\Upsilon_{\mathcal{K}}^\pm \to  \bar{f}f'$} and \mbox{$\Upsilon_{\mathcal{K}}^\pm \to W \hat{g} \hat{g}$} decays, we exploit the similarity with the SM quarkonia. For example, for the $J/\psi$ we have (see e.g. Ref.~\cite{Brambilla:2004wf})
\begin{equation}
{\Gamma(J/\psi \to \gamma g g ) \over \Gamma(J/\psi \to \gamma^* \to e^+ e^-)}\simeq \frac{8}{9}\frac{\pi^2-9}{\pi }\frac{\alpha_s^2(m_b)}{\alpha}\,.
\end{equation}
By replacing the photon with the $W$ and accounting for an extra factor $2^2$, which arises because the $W$ couples only to left-handed fermions and with coupling strength $g/\sqrt{2}$, we arrive at
\begin{equation} \label{eq:TH_BR}
{\Gamma(\Upsilon^\pm \to W^\pm \hat{g} \hat{g} ) \over \Gamma(\Upsilon^\pm \to W^{\pm\,\ast} \to \bar{f} f^\prime)} \simeq \frac{32}{9}\frac{\pi^2-9}{\pi }\frac{\hat{\alpha}_s^2(m_{\mathcal{K}^-})}{\alpha_W}\, \frac{1}{12}\,,
\end{equation}
where the factor of $1/12$ accounts for the multiplicity of the SM fermion-antifermion final states available in the decay through the off-shell $W$. The resulting branching ratio for $\Upsilon^\pm_{\mathcal{K}} \to W^\pm \hat{g} \hat{g}$ varies from $1\%$ to $5\%$ in the mass range we study. We make the assumption that the twin gluons dominantly hadronize into a single lightest glueball $\hat{G}$, which is reasonable if the glueball production can be described by a thermal process with temperature $\sim \hat{\Lambda} \ll m_{\hat{G}}$ \cite{Juknevich:2010rhj}. Notice, however, that our analysis strategy is not affected if additional glueballs are produced by the twin hadronization. Once the glueball mass is fixed, the running of $\hat{\alpha}_s$ is determined, which in turn sets the size of the wavefunction at the origin through Eq.~\eqref{eq:wavefunction} and the $\Upsilon_{\mathcal{K}}$ branching ratios via Eq.~\eqref{eq:TH_BR}. Therefore in our analysis we take $M_{\Upsilon_{\mathcal{K}}}$ and $m_{\hat{G}}$ as the two input parameters. Furthermore, for $\hat{\Lambda} \lesssim 10$ GeV we have $a_0 \hat{\Lambda} \ll 1$ for all the values of $m_{\mathcal{K}^-}$ we consider, hence it is safe to apply the Coulomb approximation.

The results are shown in Fig.~\ref{fig:THbound}. Despite the suppressed branching ratio $\mathrm{BR}(\Upsilon_{\mathcal{K}}\to W \hat{G})\sim \mathrm{few}\;\%$, this channel is competitive with $\Upsilon_{\mathcal{K}} \to \ell\nu$, because the striking combination of a prompt lepton and a DV renders the final state essentially background-free. This decay is peculiar of the UV-extended FTH model. Similarly to the case of $\lambda$-SUSY, discussed at the end of Sec.~\ref{sec:lambdasusy}, if $\hat{G}$ decays in the ID with boost factor $\gtrsim 10$ the standard reconstruction of the hadronic DV fails. The corresponding region of parameter space is hatched in black in Fig.~\ref{fig:THbound}.

As a final comment, we observe that the signature of a prompt lepton$\,$+$\,$DV can also appear in other neutral naturalness scenarios. For example, in the Folded (F-) SUSY model \cite{Burdman:2006tz} an F-stop/F-sbottom pair can be produced through DY or vector boson fusion \cite{Burdman:2008ek}. If the F-stop decays into a (likely off-shell) $W$ and an F-sbottom, the resulting F-sbottom pair forms a squirky bound state. The latter promptly annihilates into mirror glueballs, which in turn can yield displaced signatures by decaying through the Higgs portal \cite{Chacko:2015fbc}.

\section{Discussion}\label{sec:conclusion}
In this paper we have presented a new strategy to search for hidden force carriers at the LHC. These particles have suppressed direct production cross sections, due to their small couplings to the SM particles, but can be produced through mediators that carry at least some of the SM charges. We focused on the cases where the hidden force carrier $X$ is either a real scalar $\phi$ or a dark photon $\gamma_d$, and the mediators are a pair of electroweak-charged vector-like fermions $\psi^{+, 0}$. Once a $\psi^\pm \bar{\psi}^0$ pair is produced in the DY process, the strong hidden force can bind it into an electrically charged spin-$1$ bound state $\Upsilon^\pm$, which promptly annihilates into $W^\pm X$. The corresponding signatures consist of a prompt lepton originating from the $W$ boson, and a prompt or displaced $\phi \to b\bar{b}$ or $\gamma_d \to \ell \ell$ decay. We analyzed these final states in detail, estimating the LHC reach within a simplified model approach. To illustrate the impact of our results, we also applied them to two motivated example models that contain hidden forces and can yield these signatures, namely $\lambda$-SUSY and the UV-extended Fraternal Twin Higgs. The resonant signals allow for the measurement of the mass of both the bound state and the force carrier, thus yielding critical insights on the structure of the hidden sector.

For displaced $X$ decays, we proposed new searches for $(b\bar{b})$ and $(\ell \ell)$ displaced vertices, where the simultaneous presence of a hard prompt lepton stemming from the $W$ ensures efficient triggering and essentially removes all SM backgrounds. As a consequence, the reach of these searches can compete with that of the irreducible $\Upsilon^\pm \to \ell \nu$ signal even when the bound state decays to $W^\pm X$ with subdominant branching fraction. Signals of this type are especially promising for testing models of neutral naturalness. 

In the case of prompt $X$ decays, we showed that simple extensions of existing diboson searches would allow ATLAS and CMS to obtain a compelling reach. Furthermore, while the simplified analyses performed in this paper lose sensitivity when the $X$ decay products are collimated, the experimental collaborations have full capability to exploit this type of events, either by employing jet substructure variables in the $\phi \to b\bar{b}$ decay or by resolving narrowly separated leptons that originate from $\gamma_d \to \ell \ell$. We believe that our results provide further motivation for extending the array of diboson searches to include heavy resonance decays to BSM particles.


\vspace{0.5cm}
\begin{center}
{\bf Acknowledgments}
\end{center}
We thank J.~Collins, A.~De~Roeck, Y.~Jiang, B.~Shakya, C.~Verhaaren, L.-T.~Wang, and Y.~Zhao for useful discussions. LL and RZ were supported in part by the US Department of Energy grant DE-SC-000999. The work of ES has been partially supported by the DFG Cluster of Excellence 153 ``Origin and Structure of the Universe,'' by the Collaborative Research Center SFB1258 and the COST Action CA15108. YT was supported in part by the National Science Foundation under grant PHY-1315155, and by the Maryland Center for Fundamental Physics. This work was performed in part at the Aspen Center for Physics, which is supported by the National Science Foundation grant PHY-1607611. ES (YT) is grateful to the MCFP (TUM Physics Department) for hospitality in the final stages of the project.

\bibliography{./BoundState}

\begin{thebibliography}{54}%
\makeatletter
\providecommand \@ifxundefined [1]{%
 \@ifx{#1\undefined}
}%
\providecommand \@ifnum [1]{%
 \ifnum #1\expandafter \@firstoftwo
 \else \expandafter \@secondoftwo
 \fi
}%
\providecommand \@ifx [1]{%
 \ifx #1\expandafter \@firstoftwo
 \else \expandafter \@secondoftwo
 \fi
}%
\providecommand \natexlab [1]{#1}%
\providecommand \enquote  [1]{``#1''}%
\providecommand \bibnamefont  [1]{#1}%
\providecommand \bibfnamefont [1]{#1}%
\providecommand \citenamefont [1]{#1}%
\providecommand \href@noop [0]{\@secondoftwo}%
\providecommand \href [0]{\begingroup \@sanitize@url \@href}%
\providecommand \@href[1]{\@@startlink{#1}\@@href}%
\providecommand \@@href[1]{\endgroup#1\@@endlink}%
\providecommand \@sanitize@url [0]{\catcode `\\12\catcode `\$12\catcode
  `\&12\catcode `\#12\catcode `\^12\catcode `\_12\catcode `\%12\relax}%
\providecommand \@@startlink[1]{}%
\providecommand \@@endlink[0]{}%
\providecommand \url  [0]{\begingroup\@sanitize@url \@url }%
\providecommand \@url [1]{\endgroup\@href {#1}{\urlprefix }}%
\providecommand \urlprefix  [0]{URL }%
\providecommand \Eprint [0]{\href }%
\providecommand \doibase [0]{http://dx.doi.org/}%
\providecommand \selectlanguage [0]{\@gobble}%
\providecommand \bibinfo  [0]{\@secondoftwo}%
\providecommand \bibfield  [0]{\@secondoftwo}%
\providecommand \translation [1]{[#1]}%
\providecommand \BibitemOpen [0]{}%
\providecommand \bibitemStop [0]{}%
\providecommand \bibitemNoStop [0]{.\EOS\space}%
\providecommand \EOS [0]{\spacefactor3000\relax}%
\providecommand \BibitemShut  [1]{\csname bibitem#1\endcsname}%
\let\auto@bib@innerbib\@empty
\bibitem [{\citenamefont {Chacko}\ \emph {et~al.}(2006)\citenamefont {Chacko},
  \citenamefont {Goh},\ and\ \citenamefont {Harnik}}]{Chacko:2005pe}%
  \BibitemOpen
  \bibfield  {author} {\bibinfo {author} {\bibfnamefont {Z.}~\bibnamefont
  {Chacko}}, \bibinfo {author} {\bibfnamefont {H.-S.}\ \bibnamefont {Goh}}, \
  and\ \bibinfo {author} {\bibfnamefont {R.}~\bibnamefont {Harnik}},\ }\href
  {\doibase 10.1103/PhysRevLett.96.231802} {\bibfield  {journal} {\bibinfo
  {journal} {Phys. Rev. Lett.}\ }\textbf {\bibinfo {volume} {96}},\ \bibinfo
  {pages} {231802} (\bibinfo {year} {2006})},\ \Eprint
  {http://arxiv.org/abs/hep-ph/0506256} {arXiv:hep-ph/0506256} \BibitemShut
  {NoStop}%
\bibitem [{\citenamefont {Burdman}\ \emph {et~al.}(2007)\citenamefont
  {Burdman}, \citenamefont {Chacko}, \citenamefont {Goh},\ and\ \citenamefont
  {Harnik}}]{Burdman:2006tz}%
  \BibitemOpen
  \bibfield  {author} {\bibinfo {author} {\bibfnamefont {G.}~\bibnamefont
  {Burdman}}, \bibinfo {author} {\bibfnamefont {Z.}~\bibnamefont {Chacko}},
  \bibinfo {author} {\bibfnamefont {H.-S.}\ \bibnamefont {Goh}}, \ and\
  \bibinfo {author} {\bibfnamefont {R.}~\bibnamefont {Harnik}},\ }\href
  {\doibase 10.1088/1126-6708/2007/02/009} {\bibfield  {journal} {\bibinfo
  {journal} {JHEP}\ }\textbf {\bibinfo {volume} {02}},\ \bibinfo {pages} {009}
  (\bibinfo {year} {2007})},\ \Eprint {http://arxiv.org/abs/hep-ph/0609152}
  {arXiv:hep-ph/0609152} \BibitemShut {NoStop}%
\bibitem [{\citenamefont {Batra}\ \emph {et~al.}(2004)\citenamefont {Batra},
  \citenamefont {Delgado}, \citenamefont {Kaplan},\ and\ \citenamefont
  {Tait}}]{Batra:2003nj}%
  \BibitemOpen
  \bibfield  {author} {\bibinfo {author} {\bibfnamefont {P.}~\bibnamefont
  {Batra}}, \bibinfo {author} {\bibfnamefont {A.}~\bibnamefont {Delgado}},
  \bibinfo {author} {\bibfnamefont {D.~E.}\ \bibnamefont {Kaplan}}, \ and\
  \bibinfo {author} {\bibfnamefont {T.~M.~P.}\ \bibnamefont {Tait}},\ }\href
  {\doibase 10.1088/1126-6708/2004/02/043} {\bibfield  {journal} {\bibinfo
  {journal} {JHEP}\ }\textbf {\bibinfo {volume} {02}},\ \bibinfo {pages} {043}
  (\bibinfo {year} {2004})},\ \Eprint {http://arxiv.org/abs/hep-ph/0309149}
  {arXiv:hep-ph/0309149} \BibitemShut {NoStop}%
\bibitem [{\citenamefont {Maloney}\ \emph {et~al.}(2006)\citenamefont
  {Maloney}, \citenamefont {Pierce},\ and\ \citenamefont
  {Wacker}}]{Maloney:2004rc}%
  \BibitemOpen
  \bibfield  {author} {\bibinfo {author} {\bibfnamefont {A.}~\bibnamefont
  {Maloney}}, \bibinfo {author} {\bibfnamefont {A.}~\bibnamefont {Pierce}}, \
  and\ \bibinfo {author} {\bibfnamefont {J.~G.}\ \bibnamefont {Wacker}},\
  }\href {\doibase 10.1088/1126-6708/2006/06/034} {\bibfield  {journal}
  {\bibinfo  {journal} {JHEP}\ }\textbf {\bibinfo {volume} {06}},\ \bibinfo
  {pages} {034} (\bibinfo {year} {2006})},\ \Eprint
  {http://arxiv.org/abs/hep-ph/0409127} {arXiv:hep-ph/0409127} \BibitemShut
  {NoStop}%
\bibitem [{\citenamefont {Barbieri}\ \emph {et~al.}(2007)\citenamefont
  {Barbieri}, \citenamefont {Hall}, \citenamefont {Nomura},\ and\ \citenamefont
  {Rychkov}}]{Barbieri:2006bg}%
  \BibitemOpen
  \bibfield  {author} {\bibinfo {author} {\bibfnamefont {R.}~\bibnamefont
  {Barbieri}}, \bibinfo {author} {\bibfnamefont {L.~J.}\ \bibnamefont {Hall}},
  \bibinfo {author} {\bibfnamefont {Y.}~\bibnamefont {Nomura}}, \ and\ \bibinfo
  {author} {\bibfnamefont {V.~S.}\ \bibnamefont {Rychkov}},\ }\href {\doibase
  10.1103/PhysRevD.75.035007} {\bibfield  {journal} {\bibinfo  {journal} {Phys.
  Rev.}\ }\textbf {\bibinfo {volume} {D75}},\ \bibinfo {pages} {035007}
  (\bibinfo {year} {2007})},\ \Eprint {http://arxiv.org/abs/hep-ph/0607332}
  {arXiv:hep-ph/0607332} \BibitemShut {NoStop}%
\bibitem [{\citenamefont {Tulin}\ \emph {et~al.}(2013)\citenamefont {Tulin},
  \citenamefont {Yu},\ and\ \citenamefont {Zurek}}]{Tulin:2013teo}%
  \BibitemOpen
  \bibfield  {author} {\bibinfo {author} {\bibfnamefont {S.}~\bibnamefont
  {Tulin}}, \bibinfo {author} {\bibfnamefont {H.-B.}\ \bibnamefont {Yu}}, \
  and\ \bibinfo {author} {\bibfnamefont {K.~M.}\ \bibnamefont {Zurek}},\ }\href
  {\doibase 10.1103/PhysRevD.87.115007} {\bibfield  {journal} {\bibinfo
  {journal} {Phys. Rev.}\ }\textbf {\bibinfo {volume} {D87}},\ \bibinfo {pages}
  {115007} (\bibinfo {year} {2013})},\ \Eprint {http://arxiv.org/abs/1302.3898}
  {arXiv:1302.3898 [hep-ph]} \BibitemShut {NoStop}%
\bibitem [{\citenamefont {Hochberg}\ \emph {et~al.}(2014)\citenamefont
  {Hochberg}, \citenamefont {Kuflik}, \citenamefont {Volansky},\ and\
  \citenamefont {Wacker}}]{Hochberg:2014dra}%
  \BibitemOpen
  \bibfield  {author} {\bibinfo {author} {\bibfnamefont {Y.}~\bibnamefont
  {Hochberg}}, \bibinfo {author} {\bibfnamefont {E.}~\bibnamefont {Kuflik}},
  \bibinfo {author} {\bibfnamefont {T.}~\bibnamefont {Volansky}}, \ and\
  \bibinfo {author} {\bibfnamefont {J.~G.}\ \bibnamefont {Wacker}},\ }\href
  {\doibase 10.1103/PhysRevLett.113.171301} {\bibfield  {journal} {\bibinfo
  {journal} {Phys. Rev. Lett.}\ }\textbf {\bibinfo {volume} {113}},\ \bibinfo
  {pages} {171301} (\bibinfo {year} {2014})},\ \Eprint
  {http://arxiv.org/abs/1402.5143} {arXiv:1402.5143 [hep-ph]} \BibitemShut
  {NoStop}%
\bibitem [{\citenamefont {Kaplinghat}\ \emph {et~al.}(2016)\citenamefont
  {Kaplinghat}, \citenamefont {Tulin},\ and\ \citenamefont
  {Yu}}]{Kaplinghat:2015aga}%
  \BibitemOpen
  \bibfield  {author} {\bibinfo {author} {\bibfnamefont {M.}~\bibnamefont
  {Kaplinghat}}, \bibinfo {author} {\bibfnamefont {S.}~\bibnamefont {Tulin}}, \
  and\ \bibinfo {author} {\bibfnamefont {H.-B.}\ \bibnamefont {Yu}},\ }\href
  {\doibase 10.1103/PhysRevLett.116.041302} {\bibfield  {journal} {\bibinfo
  {journal} {Phys. Rev. Lett.}\ }\textbf {\bibinfo {volume} {116}},\ \bibinfo
  {pages} {041302} (\bibinfo {year} {2016})},\ \Eprint
  {http://arxiv.org/abs/1508.03339} {arXiv:1508.03339 [astro-ph.CO]}
  \BibitemShut {NoStop}%
\bibitem [{\citenamefont {{{ATLAS
  Collaboration}}}(2015{\natexlab{a}})}]{Aad:2015yza}%
  \BibitemOpen
  \bibfield  {author} {\bibinfo {author} {\bibnamefont {{{ATLAS
  Collaboration}}}},\ }\href {\doibase 10.1140/epjc/s10052-015-3474-x}
  {\bibfield  {journal} {\bibinfo  {journal} {Eur. Phys. J.}\ }\textbf
  {\bibinfo {volume} {C75}},\ \bibinfo {pages} {263} (\bibinfo {year}
  {2015}{\natexlab{a}})},\ \Eprint {http://arxiv.org/abs/1503.08089}
  {arXiv:1503.08089 [hep-ex]} \BibitemShut {NoStop}%
\bibitem [{Aab({\natexlab{a}})}]{Aaboud:2017ecz}%
  \BibitemOpen
  \href@noop {} {\emph {\bibinfo {title} {{{\rm ATLAS Collaboration}}}}},\
  \Eprint {http://arxiv.org/abs/1709.06783} {arXiv:1709.06783 [hep-ex]}
  \BibitemShut {NoStop}%
\bibitem [{\citenamefont {{{ATLAS
  Collaboration}}}(2013)}]{ATLAS-CONF-2013-015}%
  \BibitemOpen
  \bibfield  {author} {\bibinfo {author} {\bibnamefont {{{ATLAS
  Collaboration}}}},\ }\href {http://cds.cern.ch/record/1525522} {\bibfield
  {journal} {\bibinfo  {journal} {{ATLAS-CONF-2013-015}}\ } (\bibinfo {year}
  {2013})}\BibitemShut {NoStop}%
\bibitem [{\citenamefont {{{ATLAS
  Collaboration}}}(2014{\natexlab{a}})}]{Aad:2014pha}%
  \BibitemOpen
  \bibfield  {author} {\bibinfo {author} {\bibnamefont {{{ATLAS
  Collaboration}}}},\ }\href {\doibase 10.1016/j.physletb.2014.08.039}
  {\bibfield  {journal} {\bibinfo  {journal} {Phys. Lett.}\ }\textbf {\bibinfo
  {volume} {B737}},\ \bibinfo {pages} {223} (\bibinfo {year}
  {2014}{\natexlab{a}})},\ \Eprint {http://arxiv.org/abs/1406.4456}
  {arXiv:1406.4456 [hep-ex]} \BibitemShut {NoStop}%
\bibitem [{\citenamefont {Tsai}\ \emph {et~al.}(2016)\citenamefont {Tsai},
  \citenamefont {Wang},\ and\ \citenamefont {Zhao}}]{Tsai:2015ugz}%
  \BibitemOpen
  \bibfield  {author} {\bibinfo {author} {\bibfnamefont {Y.}~\bibnamefont
  {Tsai}}, \bibinfo {author} {\bibfnamefont {L.-T.}\ \bibnamefont {Wang}}, \
  and\ \bibinfo {author} {\bibfnamefont {Y.}~\bibnamefont {Zhao}},\ }\href
  {\doibase 10.1103/PhysRevD.93.035024} {\bibfield  {journal} {\bibinfo
  {journal} {Phys. Rev.}\ }\textbf {\bibinfo {volume} {D93}},\ \bibinfo {pages}
  {035024} (\bibinfo {year} {2016})},\ \Eprint
  {http://arxiv.org/abs/1511.07433} {arXiv:1511.07433 [hep-ph]} \BibitemShut
  {NoStop}%
\bibitem [{\citenamefont {Cheng}\ \emph {et~al.}(2016)\citenamefont {Cheng},
  \citenamefont {Jung}, \citenamefont {Salvioni},\ and\ \citenamefont
  {Tsai}}]{Cheng:2015buv}%
  \BibitemOpen
  \bibfield  {author} {\bibinfo {author} {\bibfnamefont {H.-C.}\ \bibnamefont
  {Cheng}}, \bibinfo {author} {\bibfnamefont {S.}~\bibnamefont {Jung}},
  \bibinfo {author} {\bibfnamefont {E.}~\bibnamefont {Salvioni}}, \ and\
  \bibinfo {author} {\bibfnamefont {Y.}~\bibnamefont {Tsai}},\ }\href {\doibase
  10.1007/JHEP03(2016)074} {\bibfield  {journal} {\bibinfo  {journal} {JHEP}\
  }\textbf {\bibinfo {volume} {03}},\ \bibinfo {pages} {074} (\bibinfo {year}
  {2016})},\ \Eprint {http://arxiv.org/abs/1512.02647} {arXiv:1512.02647
  [hep-ph]} \BibitemShut {NoStop}%
\bibitem [{\citenamefont {Cheng}\ \emph {et~al.}(2017)\citenamefont {Cheng},
  \citenamefont {Salvioni},\ and\ \citenamefont {Tsai}}]{Cheng:2016uqk}%
  \BibitemOpen
  \bibfield  {author} {\bibinfo {author} {\bibfnamefont {H.-C.}\ \bibnamefont
  {Cheng}}, \bibinfo {author} {\bibfnamefont {E.}~\bibnamefont {Salvioni}}, \
  and\ \bibinfo {author} {\bibfnamefont {Y.}~\bibnamefont {Tsai}},\ }\href@noop
  {} {\bibfield  {journal} {\bibinfo  {journal} {Phys. Rev.}\ }\textbf
  {\bibinfo {volume} {D95}},\ \bibinfo {pages} {115035} (\bibinfo {year}
  {2017})},\ \Eprint {http://arxiv.org/abs/1612.03176} {arXiv:1612.03176
  [hep-ph]} \BibitemShut {NoStop}%
\bibitem [{\citenamefont {Craig}\ \emph {et~al.}(2015)\citenamefont {Craig},
  \citenamefont {Katz}, \citenamefont {Strassler},\ and\ \citenamefont
  {Sundrum}}]{Craig:2015pha}%
  \BibitemOpen
  \bibfield  {author} {\bibinfo {author} {\bibfnamefont {N.}~\bibnamefont
  {Craig}}, \bibinfo {author} {\bibfnamefont {A.}~\bibnamefont {Katz}},
  \bibinfo {author} {\bibfnamefont {M.}~\bibnamefont {Strassler}}, \ and\
  \bibinfo {author} {\bibfnamefont {R.}~\bibnamefont {Sundrum}},\ }\href
  {\doibase 10.1007/JHEP07(2015)105} {\bibfield  {journal} {\bibinfo  {journal}
  {JHEP}\ }\textbf {\bibinfo {volume} {07}},\ \bibinfo {pages} {105} (\bibinfo
  {year} {2015})},\ \Eprint {http://arxiv.org/abs/1501.05310} {arXiv:1501.05310
  [hep-ph]} \BibitemShut {NoStop}%
\bibitem [{\citenamefont {Shepherd}\ \emph {et~al.}(2009)\citenamefont
  {Shepherd}, \citenamefont {Tait},\ and\ \citenamefont
  {Zaharijas}}]{Shepherd:2009sa}%
  \BibitemOpen
  \bibfield  {author} {\bibinfo {author} {\bibfnamefont {W.}~\bibnamefont
  {Shepherd}}, \bibinfo {author} {\bibfnamefont {T.~M.~P.}\ \bibnamefont
  {Tait}}, \ and\ \bibinfo {author} {\bibfnamefont {G.}~\bibnamefont
  {Zaharijas}},\ }\href {\doibase 10.1103/PhysRevD.79.055022} {\bibfield
  {journal} {\bibinfo  {journal} {Phys. Rev.}\ }\textbf {\bibinfo {volume}
  {D79}},\ \bibinfo {pages} {055022} (\bibinfo {year} {2009})},\ \Eprint
  {http://arxiv.org/abs/0901.2125} {arXiv:0901.2125 [hep-ph]} \BibitemShut
  {NoStop}%
\bibitem [{\citenamefont {An}\ \emph {et~al.}(2016)\citenamefont {An},
  \citenamefont {Echenard}, \citenamefont {Pospelov},\ and\ \citenamefont
  {Zhang}}]{An:2015pva}%
  \BibitemOpen
  \bibfield  {author} {\bibinfo {author} {\bibfnamefont {H.}~\bibnamefont
  {An}}, \bibinfo {author} {\bibfnamefont {B.}~\bibnamefont {Echenard}},
  \bibinfo {author} {\bibfnamefont {M.}~\bibnamefont {Pospelov}}, \ and\
  \bibinfo {author} {\bibfnamefont {Y.}~\bibnamefont {Zhang}},\ }\href
  {\doibase 10.1103/PhysRevLett.116.151801} {\bibfield  {journal} {\bibinfo
  {journal} {Phys. Rev. Lett.}\ }\textbf {\bibinfo {volume} {116}},\ \bibinfo
  {pages} {151801} (\bibinfo {year} {2016})},\ \Eprint
  {http://arxiv.org/abs/1510.05020} {arXiv:1510.05020 [hep-ph]} \BibitemShut
  {NoStop}%
\bibitem [{\citenamefont {Kats}\ and\ \citenamefont
  {Schwartz}(2010)}]{Kats:2009bv}%
  \BibitemOpen
  \bibfield  {author} {\bibinfo {author} {\bibfnamefont {Y.}~\bibnamefont
  {Kats}}\ and\ \bibinfo {author} {\bibfnamefont {M.~D.}\ \bibnamefont
  {Schwartz}},\ }\href {\doibase 10.1007/JHEP04(2010)016} {\bibfield  {journal}
  {\bibinfo  {journal} {JHEP}\ }\textbf {\bibinfo {volume} {04}},\ \bibinfo
  {pages} {016} (\bibinfo {year} {2010})},\ \Eprint
  {http://arxiv.org/abs/0912.0526} {arXiv:0912.0526 [hep-ph]} \BibitemShut
  {NoStop}%
\bibitem [{\citenamefont {Kats}\ and\ \citenamefont
  {Strassler}(2012)}]{Kats:2012ym}%
  \BibitemOpen
  \bibfield  {author} {\bibinfo {author} {\bibfnamefont {Y.}~\bibnamefont
  {Kats}}\ and\ \bibinfo {author} {\bibfnamefont {M.~J.}\ \bibnamefont
  {Strassler}},\ }\href@noop {} {\bibfield  {journal} {\bibinfo  {journal}
  {JHEP}\ }\textbf {\bibinfo {volume} {11}},\ \bibinfo {pages} {097} (\bibinfo
  {year} {2012})},\ \bibinfo {note} {[Erratum: JHEP {\bf 07}, 009 (2016)]},\
  \Eprint {http://arxiv.org/abs/1204.1119} {arXiv:1204.1119 [hep-ph]}
  \BibitemShut {NoStop}%
\bibitem [{Note1()}]{Note1}%
  \BibitemOpen
  \bibinfo {note} {Notice that in the former case we have assumed that $\alpha
  _\lambda $ does not run below the scale $m_\psi $, as it is the case in
  $\lambda $-SUSY.}\BibitemShut {Stop}%
\bibitem [{\citenamefont {{{ATLAS
  Collaboration}}}(2017{\natexlab{a}})}]{ATLAS-CONF-2017-017}%
  \BibitemOpen
  \bibfield  {author} {\bibinfo {author} {\bibnamefont {{{ATLAS
  Collaboration}}}},\ }\href {https://cds.cern.ch/record/2258131} {\bibfield
  {journal} {\bibinfo  {journal} {{{ATLAS-CONF-2017-017}}}\ } (\bibinfo {year}
  {2017}{\natexlab{a}})}\BibitemShut {NoStop}%
\bibitem [{\citenamefont {{{CMS
  Collaboration}}}(2015{\natexlab{a}})}]{CMS:2014gxa}%
  \BibitemOpen
  \bibfield  {author} {\bibinfo {author} {\bibnamefont {{{CMS
  Collaboration}}}},\ }\href@noop {} {\bibfield  {journal} {\bibinfo  {journal}
  {JHEP}\ }\textbf {\bibinfo {volume} {01}},\ \bibinfo {pages} {096} (\bibinfo
  {year} {2015}{\natexlab{a}})},\ \Eprint {http://arxiv.org/abs/1411.6006}
  {arXiv:1411.6006 [hep-ex]} \BibitemShut {NoStop}%
\bibitem [{\citenamefont {Alloul}\ \emph {et~al.}(2014)\citenamefont {Alloul},
  \citenamefont {Christensen}, \citenamefont {Degrande}, \citenamefont {Duhr},\
  and\ \citenamefont {Fuks}}]{Alloul:2013bka}%
  \BibitemOpen
  \bibfield  {author} {\bibinfo {author} {\bibfnamefont {A.}~\bibnamefont
  {Alloul}}, \bibinfo {author} {\bibfnamefont {N.~D.}\ \bibnamefont
  {Christensen}}, \bibinfo {author} {\bibfnamefont {C.}~\bibnamefont
  {Degrande}}, \bibinfo {author} {\bibfnamefont {C.}~\bibnamefont {Duhr}}, \
  and\ \bibinfo {author} {\bibfnamefont {B.}~\bibnamefont {Fuks}},\ }\href
  {\doibase 10.1016/j.cpc.2014.04.012} {\bibfield  {journal} {\bibinfo
  {journal} {Comput. Phys. Commun.}\ }\textbf {\bibinfo {volume} {185}},\
  \bibinfo {pages} {2250} (\bibinfo {year} {2014})},\ \Eprint
  {http://arxiv.org/abs/1310.1921} {arXiv:1310.1921 [hep-ph]} \BibitemShut
  {NoStop}%
\bibitem [{\citenamefont {Alwall}\ \emph {et~al.}(2014)\citenamefont {Alwall}
  \emph {et~al.}}]{Alwall:2014hca}%
  \BibitemOpen
  \bibfield  {author} {\bibinfo {author} {\bibfnamefont {J.}~\bibnamefont
  {Alwall}} \emph {et~al.},\ }\href {\doibase 10.1007/JHEP07(2014)079}
  {\bibfield  {journal} {\bibinfo  {journal} {JHEP}\ }\textbf {\bibinfo
  {volume} {07}},\ \bibinfo {pages} {079} (\bibinfo {year} {2014})},\ \Eprint
  {http://arxiv.org/abs/1405.0301} {arXiv:1405.0301 [hep-ph]} \BibitemShut
  {NoStop}%
\bibitem [{\citenamefont {Sjostrand}\ \emph {et~al.}(2006)\citenamefont
  {Sjostrand}, \citenamefont {Mrenna},\ and\ \citenamefont
  {Skands}}]{Sjostrand:2006za}%
  \BibitemOpen
  \bibfield  {author} {\bibinfo {author} {\bibfnamefont {T.}~\bibnamefont
  {Sjostrand}}, \bibinfo {author} {\bibfnamefont {S.}~\bibnamefont {Mrenna}}, \
  and\ \bibinfo {author} {\bibfnamefont {P.~Z.}\ \bibnamefont {Skands}},\
  }\href {\doibase 10.1088/1126-6708/2006/05/026} {\bibfield  {journal}
  {\bibinfo  {journal} {JHEP}\ }\textbf {\bibinfo {volume} {05}},\ \bibinfo
  {pages} {026} (\bibinfo {year} {2006})},\ \Eprint
  {http://arxiv.org/abs/hep-ph/0603175} {arXiv:hep-ph/0603175} \BibitemShut
  {NoStop}%
\bibitem [{\citenamefont {de~Favereau}\ \emph {et~al.}(2014)\citenamefont
  {de~Favereau} \emph {et~al.}}]{deFavereau:2013fsa}%
  \BibitemOpen
  \bibfield  {author} {\bibinfo {author} {\bibfnamefont {J.}~\bibnamefont
  {de~Favereau}} \emph {et~al.},\ }\href {\doibase 10.1007/JHEP02(2014)057}
  {\bibfield  {journal} {\bibinfo  {journal} {JHEP}\ }\textbf {\bibinfo
  {volume} {02}},\ \bibinfo {pages} {057} (\bibinfo {year} {2014})},\ \Eprint
  {http://arxiv.org/abs/1307.6346} {arXiv:1307.6346 [hep-ex]} \BibitemShut
  {NoStop}%
\bibitem [{\citenamefont {Avetisyan}\ \emph {et~al.}(2013)\citenamefont
  {Avetisyan} \emph {et~al.}}]{Avetisyan:2013onh}%
  \BibitemOpen
  \bibfield  {author} {\bibinfo {author} {\bibfnamefont {A.}~\bibnamefont
  {Avetisyan}} \emph {et~al.},\ }in\ \href
  {http://lss.fnal.gov/archive/test-fn/0000/fermilab-fn-0965-t.pdf} {\emph
  {\bibinfo {booktitle} {{Proceedings, 2013 Community Summer Study: Snowmass on
  the Mississippi: Minneapolis, MN, USA}}}}\ (\bibinfo {year} {2013})\ \Eprint
  {http://arxiv.org/abs/1308.1636} {, arXiv:1308.1636 [hep-ex]} \BibitemShut
  {NoStop}%
\bibitem [{\citenamefont {Anderson}\ \emph {et~al.}(2013)\citenamefont
  {Anderson} \emph {et~al.}}]{Anderson:2013kxz}%
  \BibitemOpen
  \bibfield  {author} {\bibinfo {author} {\bibfnamefont {J.}~\bibnamefont
  {Anderson}} \emph {et~al.},\ }in\ \href
  {https://inspirehep.net/record/1252716/files/arXiv:1309.1057.pdf} {\emph
  {\bibinfo {booktitle} {{Proceedings, 2013 Community Summer Study: Snowmass on
  the Mississippi: Minneapolis, MN, USA}}}}\ (\bibinfo {year} {2013})\ \Eprint
  {http://arxiv.org/abs/1309.1057} {, arXiv:1309.1057 [hep-ex]} \BibitemShut
  {NoStop}%
\bibitem [{\citenamefont {{{ATLAS
  Collaboration}}}(2014{\natexlab{b}})}]{ATLAS:2014jfa}%
  \BibitemOpen
  \bibfield  {author} {\bibinfo {author} {\bibnamefont {{{ATLAS
  Collaboration}}}},\ }\href {https://cds.cern.ch/record/1664335} {\bibfield
  {journal} {\bibinfo  {journal} {ATLAS-CONF-2014-004}\ } (\bibinfo {year}
  {2014}{\natexlab{b}})}\BibitemShut {NoStop}%
\bibitem [{Note2()}]{Note2}%
  \BibitemOpen
  \bibinfo {note} {The $W$ transverse mass is defined as $m_T^W = \protect
  \sqrt {2[{\mathchoice {{\setbox \z@ \hbox {$\mathsurround \z@ \displaystyle
  E$}\setbox \tw@ \hbox {$\mathsurround \z@ \displaystyle /$}\dimen 4\wd \z@
  \dimen@ \ht \tw@ \advance \dimen@ -\dp \tw@ \advance \dimen@ -\ht \z@
  \advance \dimen@ \dp \z@ \divide \dimen@ \tw@ \advance \dimen@ -0\ht \tw@
  \advance \dimen@ -0\dp \tw@ \dimen@ii 0\wd \z@ \raise -\dimen@ \hbox to\dimen
  4{\hss \kern \dimen@ii \box \tw@ \kern -\dimen@ii \hss }\hbox to\z@ {\hss
  \hbox to\dimen 4{\hss \box \z@ \hss }}}}{{\setbox \z@ \hbox {$\mathsurround
  \z@ \textstyle E$}\setbox \tw@ \hbox {$\mathsurround \z@ \textstyle /$}\dimen
  4\wd \z@ \dimen@ \ht \tw@ \advance \dimen@ -\dp \tw@ \advance \dimen@ -\ht
  \z@ \advance \dimen@ \dp \z@ \divide \dimen@ \tw@ \advance \dimen@ -0\ht \tw@
  \advance \dimen@ -0\dp \tw@ \dimen@ii 0\wd \z@ \raise -\dimen@ \hbox to\dimen
  4{\hss \kern \dimen@ii \box \tw@ \kern -\dimen@ii \hss }\hbox to\z@ {\hss
  \hbox to\dimen 4{\hss \box \z@ \hss }}}}{{\setbox \z@ \hbox {$\mathsurround
  \z@ \scriptstyle E$}\setbox \tw@ \hbox {$\mathsurround \z@ \scriptstyle
  /$}\dimen 4\wd \z@ \dimen@ \ht \tw@ \advance \dimen@ -\dp \tw@ \advance
  \dimen@ -\ht \z@ \advance \dimen@ \dp \z@ \divide \dimen@ \tw@ \advance
  \dimen@ -0\ht \tw@ \advance \dimen@ -0\dp \tw@ \dimen@ii 0\wd \z@ \raise
  -\dimen@ \hbox to\dimen 4{\hss \kern \dimen@ii \box \tw@ \kern -\dimen@ii
  \hss }\hbox to\z@ {\hss \hbox to\dimen 4{\hss \box \z@ \hss }}}}{{\setbox \z@
  \hbox {$\mathsurround \z@ \scriptscriptstyle E$}\setbox \tw@ \hbox
  {$\mathsurround \z@ \scriptscriptstyle /$}\dimen 4\wd \z@ \dimen@ \ht \tw@
  \advance \dimen@ -\dp \tw@ \advance \dimen@ -\ht \z@ \advance \dimen@ \dp \z@
  \divide \dimen@ \tw@ \advance \dimen@ -0\ht \tw@ \advance \dimen@ -0\dp \tw@
  \dimen@ii 0\wd \z@ \raise -\dimen@ \hbox to\dimen 4{\hss \kern \dimen@ii \box
  \tw@ \kern -\dimen@ii \hss }\hbox to\z@ {\hss \hbox to\dimen 4{\hss \box \z@
  \hss }}}}}_T p_{T}^{\ell } (1-\protect \qopname \relax o{cos}\Delta \phi
  )]}$, where $\Delta \phi $ is the azimuthal separation between the MET vector
  and the lepton momentum.}\BibitemShut {Stop}%
\bibitem [{Note3()}]{Note3}%
  \BibitemOpen
  \bibinfo {note} {Following Ref.~\cite {Aad:2015yza}, if the quadratic
  equation has two real solutions for $p^{z}_\nu $, then we take the one with
  smaller absolute value. If the solutions are complex, we take the real
  part.}\BibitemShut {Stop}%
\bibitem [{\citenamefont {{{CMS Collaboration}}}(2014)}]{CMS:2014eda}%
  \BibitemOpen
  \bibfield  {author} {\bibinfo {author} {\bibnamefont {{{CMS
  Collaboration}}}},\ }\href {http://cds.cern.ch/record/1748425} {\bibfield
  {journal} {\bibinfo  {journal} {CMS-PAS-HIG-14-013}\ } (\bibinfo {year}
  {2014})}\BibitemShut {NoStop}%
\bibitem [{Note4()}]{Note4}%
  \BibitemOpen
  \bibinfo {note} {Notice also that if $m_\phi < m_h/2 = 62.5\protect \tmspace
  +\thickmuskip {.2777em}\protect \mathrm {GeV}$, important, albeit
  model-dependent, constraints can arise from the $h \to \phi \phi $
  decay.}\BibitemShut {Stop}%
\bibitem [{\citenamefont {Butterworth}\ \emph {et~al.}(2008)\citenamefont
  {Butterworth}, \citenamefont {Davison}, \citenamefont {Rubin},\ and\
  \citenamefont {Salam}}]{Butterworth:2008iy}%
  \BibitemOpen
  \bibfield  {author} {\bibinfo {author} {\bibfnamefont {J.~M.}\ \bibnamefont
  {Butterworth}}, \bibinfo {author} {\bibfnamefont {A.~R.}\ \bibnamefont
  {Davison}}, \bibinfo {author} {\bibfnamefont {M.}~\bibnamefont {Rubin}}, \
  and\ \bibinfo {author} {\bibfnamefont {G.~P.}\ \bibnamefont {Salam}},\ }\href
  {\doibase 10.1103/PhysRevLett.100.242001} {\bibfield  {journal} {\bibinfo
  {journal} {Phys. Rev. Lett.}\ }\textbf {\bibinfo {volume} {100}},\ \bibinfo
  {pages} {242001} (\bibinfo {year} {2008})},\ \Eprint
  {http://arxiv.org/abs/0802.2470} {arXiv:0802.2470 [hep-ph]} \BibitemShut
  {NoStop}%
\bibitem [{\citenamefont {{{ATLAS
  Collaboration}}}(2017{\natexlab{b}})}]{Aaboud:2017ahz}%
  \BibitemOpen
  \bibfield  {author} {\bibinfo {author} {\bibnamefont {{{ATLAS
  Collaboration}}}},\ }\href {\doibase 10.1016/j.physletb.2017.09.066}
  {\bibfield  {journal} {\bibinfo  {journal} {Phys. Lett.}\ }\textbf {\bibinfo
  {volume} {B774}},\ \bibinfo {pages} {494} (\bibinfo {year}
  {2017}{\natexlab{b}})},\ \Eprint {http://arxiv.org/abs/1707.06958}
  {arXiv:1707.06958 [hep-ex]} \BibitemShut {NoStop}%
\bibitem [{\citenamefont {{{ATLAS
  Collaboration}}}(2015{\natexlab{b}})}]{Aad:2015asa}%
  \BibitemOpen
  \bibfield  {author} {\bibinfo {author} {\bibnamefont {{{ATLAS
  Collaboration}}}},\ }\href {\doibase 10.1016/j.physletb.2015.02.015}
  {\bibfield  {journal} {\bibinfo  {journal} {Phys. Lett.}\ }\textbf {\bibinfo
  {volume} {B743}},\ \bibinfo {pages} {15} (\bibinfo {year}
  {2015}{\natexlab{b}})},\ \Eprint {http://arxiv.org/abs/1501.04020}
  {arXiv:1501.04020 [hep-ex]} \BibitemShut {NoStop}%
\bibitem [{\citenamefont {{{ATLAS
  Collaboration}}}(2015{\natexlab{c}})}]{Aad:2015uaa}%
  \BibitemOpen
  \bibfield  {author} {\bibinfo {author} {\bibnamefont {{{ATLAS
  Collaboration}}}},\ }\href@noop {} {\bibfield  {journal} {\bibinfo  {journal}
  {Phys. Rev.}\ }\textbf {\bibinfo {volume} {D92}},\ \bibinfo {pages} {012010}
  (\bibinfo {year} {2015}{\natexlab{c}})},\ \Eprint
  {http://arxiv.org/abs/1504.03634} {arXiv:1504.03634 [hep-ex]} \BibitemShut
  {NoStop}%
\bibitem [{Note5()}]{Note5}%
  \BibitemOpen
  \bibinfo {note} {In the analysis of the twin bottomonium signals of
  Ref.~\cite {Cheng:2015buv}, the efficiency was very conservatively assumed to
  be $10\%$ also in the HCAL$+$MS. Based on the results of Ref.~\cite
  {Aad:2015uaa}, we believe $40\%$ to be closer to the actual experimental
  performance.}\BibitemShut {Stop}%
\bibitem [{\citenamefont {{{CMS
  Collaboration}}}(2015{\natexlab{b}})}]{CMS:2014hka}%
  \BibitemOpen
  \bibfield  {author} {\bibinfo {author} {\bibnamefont {{{CMS
  Collaboration}}}},\ }\href {\doibase 10.1103/PhysRevD.91.052012} {\bibfield
  {journal} {\bibinfo  {journal} {Phys. Rev.}\ }\textbf {\bibinfo {volume}
  {D91}},\ \bibinfo {pages} {052012} (\bibinfo {year} {2015}{\natexlab{b}})},\
  \Eprint {http://arxiv.org/abs/1411.6977} {arXiv:1411.6977 [hep-ex]}
  \BibitemShut {NoStop}%
\bibitem [{\citenamefont {{{ATLAS
  Collaboration}}}(2016{\natexlab{a}})}]{ATLAS-CONF-2016-042}%
  \BibitemOpen
  \bibfield  {author} {\bibinfo {author} {\bibnamefont {{{ATLAS
  Collaboration}}}},\ }\href {http://cds.cern.ch/record/2206083} {\bibfield
  {journal} {\bibinfo  {journal} {{{ATLAS-CONF-2016-042}}}\ } (\bibinfo {year}
  {2016}{\natexlab{a}})}\BibitemShut {NoStop}%
\bibitem [{Aab({\natexlab{b}})}]{Aaboud:2017efa}%
  \BibitemOpen
  \href@noop {} {\emph {\bibinfo {title} {{{\rm ATLAS Collaboration}}}}},\
  \Eprint {http://arxiv.org/abs/1706.04786} {arXiv:1706.04786 [hep-ex]}
  \BibitemShut {NoStop}%
\bibitem [{\citenamefont {{{ATLAS
  Collaboration}}}(2016{\natexlab{b}})}]{ATLAS:2016ecs}%
  \BibitemOpen
  \bibfield  {author} {\bibinfo {author} {\bibnamefont {{{ATLAS
  Collaboration}}}},\ }\href {https://cds.cern.ch/record/2206177} {\bibfield
  {journal} {\bibinfo  {journal} {ATLAS-CONF-2016-061}\ } (\bibinfo {year}
  {2016}{\natexlab{b}})}\BibitemShut {NoStop}%
\bibitem [{\citenamefont {Hall}\ \emph {et~al.}(2012)\citenamefont {Hall},
  \citenamefont {Pinner},\ and\ \citenamefont {Ruderman}}]{Hall:2011aa}%
  \BibitemOpen
  \bibfield  {author} {\bibinfo {author} {\bibfnamefont {L.~J.}\ \bibnamefont
  {Hall}}, \bibinfo {author} {\bibfnamefont {D.}~\bibnamefont {Pinner}}, \ and\
  \bibinfo {author} {\bibfnamefont {J.~T.}\ \bibnamefont {Ruderman}},\ }\href
  {\doibase 10.1007/JHEP04(2012)131} {\bibfield  {journal} {\bibinfo  {journal}
  {JHEP}\ }\textbf {\bibinfo {volume} {04}},\ \bibinfo {pages} {131} (\bibinfo
  {year} {2012})},\ \Eprint {http://arxiv.org/abs/1112.2703} {arXiv:1112.2703
  [hep-ph]} \BibitemShut {NoStop}%
\bibitem [{Note6()}]{Note6}%
  \BibitemOpen
  \bibinfo {note} {The first term in the superpotential is normalized such that
  the coupling of the physical scalar singlet $s$ to the Higgsinos is simply
  $\lambda $.}\BibitemShut {Stop}%
\bibitem [{\citenamefont {Farina}\ \emph {et~al.}(2014)\citenamefont {Farina},
  \citenamefont {Perelstein},\ and\ \citenamefont {Shakya}}]{Farina:2013fsa}%
  \BibitemOpen
  \bibfield  {author} {\bibinfo {author} {\bibfnamefont {M.}~\bibnamefont
  {Farina}}, \bibinfo {author} {\bibfnamefont {M.}~\bibnamefont {Perelstein}},
  \ and\ \bibinfo {author} {\bibfnamefont {B.}~\bibnamefont {Shakya}},\ }\href
  {\doibase 10.1007/JHEP04(2014)108} {\bibfield  {journal} {\bibinfo  {journal}
  {JHEP}\ }\textbf {\bibinfo {volume} {04}},\ \bibinfo {pages} {108} (\bibinfo
  {year} {2014})},\ \Eprint {http://arxiv.org/abs/1310.0459} {arXiv:1310.0459
  [hep-ph]} \BibitemShut {NoStop}%
\bibitem [{\citenamefont {Arkani-Hamed}\ \emph {et~al.}(2016)\citenamefont
  {Arkani-Hamed}, \citenamefont {Han}, \citenamefont {Mangano},\ and\
  \citenamefont {Wang}}]{Arkani-Hamed:2015vfh}%
  \BibitemOpen
  \bibfield  {author} {\bibinfo {author} {\bibfnamefont {N.}~\bibnamefont
  {Arkani-Hamed}}, \bibinfo {author} {\bibfnamefont {T.}~\bibnamefont {Han}},
  \bibinfo {author} {\bibfnamefont {M.}~\bibnamefont {Mangano}}, \ and\
  \bibinfo {author} {\bibfnamefont {L.-T.}\ \bibnamefont {Wang}},\ }\href
  {\doibase 10.1016/j.physrep.2016.07.004} {\bibfield  {journal} {\bibinfo
  {journal} {Phys. Rept.}\ }\textbf {\bibinfo {volume} {652}},\ \bibinfo
  {pages} {1} (\bibinfo {year} {2016})},\ \Eprint
  {http://arxiv.org/abs/1511.06495} {arXiv:1511.06495 [hep-ph]} \BibitemShut
  {NoStop}%
\bibitem [{\citenamefont {Franceschini}\ and\ \citenamefont
  {Gori}(2011)}]{Franceschini:2010qz}%
  \BibitemOpen
  \bibfield  {author} {\bibinfo {author} {\bibfnamefont {R.}~\bibnamefont
  {Franceschini}}\ and\ \bibinfo {author} {\bibfnamefont {S.}~\bibnamefont
  {Gori}},\ }\href {\doibase 10.1007/JHEP05(2011)084} {\bibfield  {journal}
  {\bibinfo  {journal} {JHEP}\ }\textbf {\bibinfo {volume} {05}},\ \bibinfo
  {pages} {084} (\bibinfo {year} {2011})},\ \Eprint
  {http://arxiv.org/abs/1005.1070} {arXiv:1005.1070 [hep-ph]} \BibitemShut
  {NoStop}%
\bibitem [{Note7()}]{Note7}%
  \BibitemOpen
  \bibinfo {note} {Precisely, $q_{\protect \rm rms} = (\protect \sqrt {3}
  \protect \tmspace +\thinmuskip {.1667em}a_0)^{-1}$, where $a_0 = 2/(C_F
  \protect \mathaccentV {hat}05E{\alpha }_s (q_{\protect \rm rms}) m_{\protect
  \mathcal {K}^-})$ is the Bohr radius and $C_F = 4/3$ \cite
  {Kats:2012ym}.}\BibitemShut {Stop}%
\bibitem [{\citenamefont {Chen}\ \emph {et~al.}(2006)\citenamefont {Chen} \emph
  {et~al.}}]{Chen:2005mg}%
  \BibitemOpen
  \bibfield  {author} {\bibinfo {author} {\bibfnamefont {Y.}~\bibnamefont
  {Chen}} \emph {et~al.},\ }\href {\doibase 10.1103/PhysRevD.73.014516}
  {\bibfield  {journal} {\bibinfo  {journal} {Phys. Rev.}\ }\textbf {\bibinfo
  {volume} {D73}},\ \bibinfo {pages} {014516} (\bibinfo {year} {2006})},\
  \Eprint {http://arxiv.org/abs/hep-lat/0510074} {arXiv:hep-lat/0510074}
  \BibitemShut {NoStop}%
\bibitem [{\citenamefont {Brambilla}\ \emph {et~al.}()\citenamefont {Brambilla}
  \emph {et~al.}}]{Brambilla:2004wf}%
  \BibitemOpen
  \bibfield  {author} {\bibinfo {author} {\bibfnamefont {N.}~\bibnamefont
  {Brambilla}} \emph {et~al.} (\bibinfo {collaboration} {Quarkonium Working
  Group}),\ }\href@noop {} {\ }\Eprint {http://arxiv.org/abs/hep-ph/0412158}
  {arXiv:hep-ph/0412158} \BibitemShut {NoStop}%
\bibitem [{\citenamefont {Juknevich}(2010)}]{Juknevich:2010rhj}%
  \BibitemOpen
  \bibfield  {author} {\bibinfo {author} {\bibfnamefont {J.~E.}\ \bibnamefont
  {Juknevich}},\ }\href {\doibase 10.7282/T34F1QHM} {Ph.D. thesis},\ \bibinfo
  {school} {Rutgers U.} (\bibinfo {year} {2010})\BibitemShut {NoStop}%
\bibitem [{\citenamefont {Burdman}\ \emph {et~al.}(2008)\citenamefont
  {Burdman}, \citenamefont {Chacko}, \citenamefont {Goh}, \citenamefont
  {Harnik},\ and\ \citenamefont {Krenke}}]{Burdman:2008ek}%
  \BibitemOpen
  \bibfield  {author} {\bibinfo {author} {\bibfnamefont {G.}~\bibnamefont
  {Burdman}}, \bibinfo {author} {\bibfnamefont {Z.}~\bibnamefont {Chacko}},
  \bibinfo {author} {\bibfnamefont {H.-S.}\ \bibnamefont {Goh}}, \bibinfo
  {author} {\bibfnamefont {R.}~\bibnamefont {Harnik}}, \ and\ \bibinfo {author}
  {\bibfnamefont {C.~A.}\ \bibnamefont {Krenke}},\ }\href {\doibase
  10.1103/PhysRevD.78.075028} {\bibfield  {journal} {\bibinfo  {journal} {Phys.
  Rev.}\ }\textbf {\bibinfo {volume} {D78}},\ \bibinfo {pages} {075028}
  (\bibinfo {year} {2008})},\ \Eprint {http://arxiv.org/abs/0805.4667}
  {arXiv:0805.4667 [hep-ph]} \BibitemShut {NoStop}%
\bibitem [{\citenamefont {Chacko}\ \emph {et~al.}(2016)\citenamefont {Chacko},
  \citenamefont {Curtin},\ and\ \citenamefont {Verhaaren}}]{Chacko:2015fbc}%
  \BibitemOpen
  \bibfield  {author} {\bibinfo {author} {\bibfnamefont {Z.}~\bibnamefont
  {Chacko}}, \bibinfo {author} {\bibfnamefont {D.}~\bibnamefont {Curtin}}, \
  and\ \bibinfo {author} {\bibfnamefont {C.~B.}\ \bibnamefont {Verhaaren}},\
  }\href {\doibase 10.1103/PhysRevD.94.011504} {\bibfield  {journal} {\bibinfo
  {journal} {Phys. Rev.}\ }\textbf {\bibinfo {volume} {D94}},\ \bibinfo {pages}
  {011504} (\bibinfo {year} {2016})},\ \Eprint
  {http://arxiv.org/abs/1512.05782} {arXiv:1512.05782 [hep-ph]} \BibitemShut
  {NoStop}%
\end{thebibliography}%

\end{document}